\documentclass[11pt,a4paper]{article}
\usepackage{jcappub}
\usepackage{float}
  
\title{Model Selection Using Cosmic Chronometers with Gaussian Processes}
\author[1,3]{Fulvio~Melia%
\note{John Woodruff Simpson Fellow.}}
\affiliation{$^3$Department of Physics, the Applied Math Program, and Steward Observatory, \\
              \null\hskip 0.05in The University of Arizona, Tucson, AZ 85721}
\emailAdd{fmelia@email.arizona.edu; manojy@email.arizona.edu}
\author[2]{and Manoj~K. Yennapureddy}
\affiliation{$^2$Department of Physics, 
              The University of Arizona,
              Tucson, AZ 85721}

\abstract{The use of Gaussian Processes with a measurement of the cosmic expansion rate
based solely on the observation of cosmic chronometers provides a completely
cosmology-independent reconstruction of the Hubble constant $H(z)$ suitable for
testing different models. The corresponding dispersion $\sigma_{H}$
is smaller than $\sim 9\%$ over the entire redshift range ($0\lesssim z\lesssim 2$)
of the observations, rivaling many kinds of cosmological measurements available
today. We use the reconstructed $H(z)$ function to test six different cosmologies,
and show that it favours the $R_{\rm h}=ct$ universe, which has only one free
parameter (i.e., $H_0$) over other models, including {\it Planck} $\Lambda$CDM.
The parameters of the standard model may be re-optimized to improve the fits to
the reconstructed $H(z)$ function, but the results have smaller $p$-values
than one finds with $R_{\rm h}=ct$.
}

\begin{document}
\maketitle
 
 \flushbottom

\section{Introduction}
Two recent developments have made it possible for us to measure the Hubble constant
$H(z)$ without having to assume any particular model, thereby providing a truly
model-independent probe of the expansion dynamics. The first of these is the development
of a technique used to measure the differential ages of adjacent galaxies out to a redshift
$z\sim 2$ \cite{1,2}. The second is the introduction of Gaussian Processes to the analysis
of the variables, such as $H(z)$, allowing us to reconstruct their functional form without having to
assume any {\it a priori} parametric dependence on redshift or other theoretical constraints
\cite{3,4,5}.

Galaxies evolving passively over a time much longer than their age difference allow us to measure
the expansion rate $H(z)$ solely as a function of the redshift-time derivative $dz/dt$. These
ages are inferred from the observed 4,000 \AA~break in the passively evolving spectra, based on
our understanding that, for old stars, this break is due to metal absorption lines whose amplitude 
scales linearly with stellar age. So when the metallicity of these stars is known, the age difference 
of two adjacent galaxies is proportional to the difference of their 4,000 \AA~amplitudes \cite{2}.  
In previous applications \cite{2,6,7,8}, these cosmic chronometers (as they are called) have been used
successfully to compare the predictions of various cosmological models, such as the standard
model $\Lambda$CDM and another Friedmann-Robertson-Walker cosmology known as the $R_{\rm h}=ct$ 
universe \cite{9,10,11}. Quite surprisingly, a constant expansion rate is preferred when 
the Hubble constant $H(z)$ is measured using cosmic chronometers on their own, without the 
bias introduced via the inclusion of other data at low $z$, such as the measurement
of $H_0$ using Cepheid variables, whose peculiar velocities associated with the
influence of a local ``Hubble Bubble" are comparable to those in the Hubble flow \cite{12}.
First attempts at identifying the distance beyond which the Hubble flow dominates
noticeably over local, peculiar velocities yielded an estimate of $\sim 80h^{-1}$ Mpc,
where $h$ is the Hubble constant scaled to $100$ km s$^{-1}$ Mpc$^{-1}$ \cite{13}.  
More recent studies \cite{14} of the local expansion rate have found a significant 
local under-density that persists out to $\sim300$ Mpc, corresponding to a redshift 
$z\sim 0.07$ (see also ref.~\cite{15}). This effect may partly be the source of 
tension between cosmological parameters optimized at low redshifts compared 
to the values obtained by {\it Planck}. Instead, the expansion rate measured 
with cosmic chronometers appears to favour the $R_{\rm h}=ct$ model, in which the 
Universe expands at a constant rate.

This situation clearly calls for more in-depth analysis of the cosmic chronometer
measurements, preferably using several different approaches for model selection. Our
goal in this paper is to follow an alternative means of using cosmic chronometers to
comparatively test these cosmologies. The use of Gaussian Processes to reconstruct
$H(z)$ avoids the need of ``fitting" the data with pre-determined parametric functions.
This non-parametric technique for reconstructing the expansion history is a fully
Bayesian approach for smoothing data. The procedure results not only in a truly
model-independent determination of $H(z)$ as a function of $z$, but the associated
errors reconstructed along with the function itself strike a balance between very
smooth and rapidly oscillating variations (see, e.g., ref.~\cite{5}). Having
said this, an important caveat to consider along with the results presented
later in this paper is that the use of Gaussian Processes necessitates the
adoption of two hyperparameters whose values are not known a priori. As discussed
in greater detail in \S~2 below, the common approach is to train them by
maximizing the likelihood that the reconstruction matches the measured values
at the data points themselves. Nonetheless, this choice of parameters affects
the smoothness of the function and its errors.

In \S~2 of this paper, we describe the calculation of $H(z)$ in more detail and
use Gaussian Processes to ``measure" this quantity out to a redshift $\sim 2$.
We then introduce the six models we will test using this metric, and carry out
the comparative analysis in \S~3. Finally, we will discuss our results and provide 
our conclusions in \S~4.

\section{Reconstructing $H(z)$}
In recent years, several kinds of measurement of $H(z)$ have been used
to optimize the parameters in $\Lambda$CDM. In some cases \cite{7,8},
they have also been used to compare the predictions of the standard
model with those of another Friedmann-Robertson-Walker (FRW) cosmology
known as the $R_{\rm h}=ct$ universe \cite{9,10,11,16,17}. But
one must be wary of combining measurements of $H(z)$ using different
techniques because not all of these approaches are truly model
independent. For example, measurements based on the identification
of baryon acoustic oscillations (BAO) and the Alcock-Paczy\'nski
distortion from galaxy clustering depend on how `standard rulers'
evolve with redshift, rather than how cosmic time changes with $z$
(see, e.g., ref.~\cite{18}). The values of $H(z)$ measured using
these different approaches are sometimes merged together to form
an overall $H(z)$ versus $z$ diagram, but the BAO approach
must necessarily adopt a particular cosmology and is therefore
model-dependent. Up to this point, observations based on the
cosmic chronometer idea are the only ones we are aware of that 
do not rely on the assumption of a particular cosmology, so they 
are suitable for testing and comparing different models. As we
explain below, however, there is a lingering concern regarding 
whether or not the assumption of a single, dominant bout of star 
formation is valid in these galaxies, and the resolution to
this question may depend on the expansion dynamics associated
with each particular model. 

\begin{figure}
\begin{center}
\includegraphics[width=5.3in]{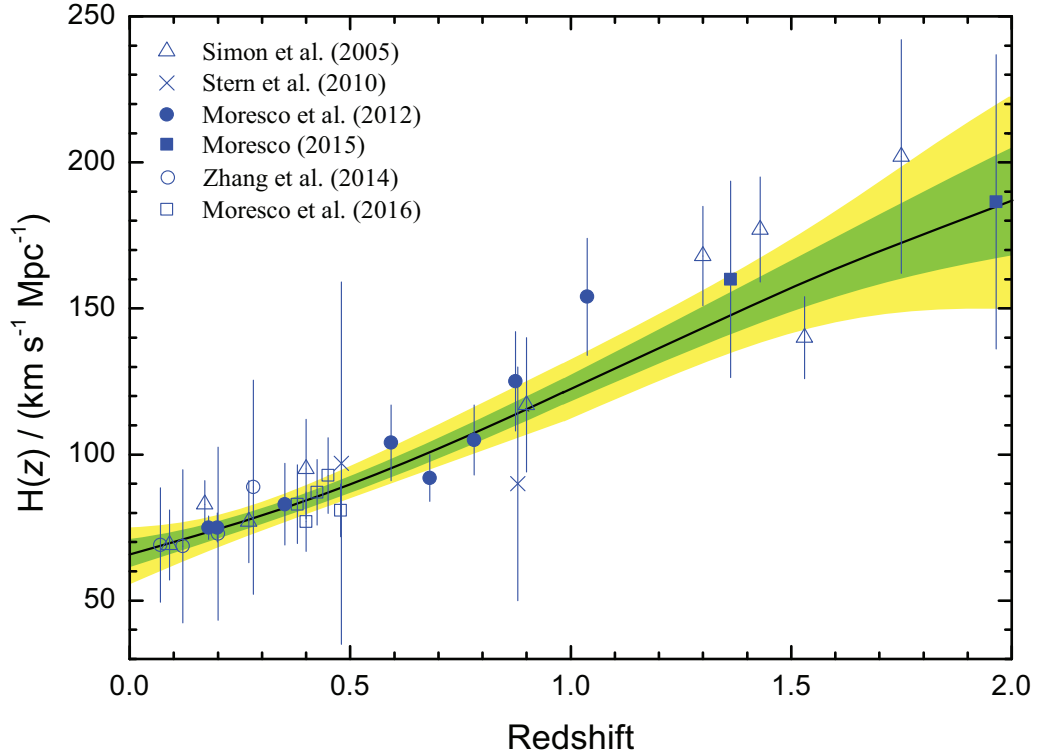}
\end{center}
\caption{Thirty model-independent measurements of $H(z)$ versus $z$ with error
bars \cite{2,6,20,21,22,23}. Note that these are total errors, which include both
statistical and systematic contributions (see, e.g., refs.~\cite{24,25}). Also shown here 
is the function $H(z)$ reconstructed with Gaussian Processes (solid black). The green and 
yellow shaded regions represent the $1\sigma$ and $2\sigma$ confidence regions, respectively, 
of the reconstruction.}
\label{figure1}
\end{figure}

\subsection{Cosmic Chronometers}
The cosmic chronometer approach is based on the notion that the
expansion rate $H(z)$ may be measured using solely the redshift-time
derivative $dz/dt$ between galaxies evolving passively over a time much
longer than their age difference \cite{1}. These massive
($\gtrsim 10^{11}\;M_\odot$) early-type galaxies are thought to have
formed $\gtrsim 90\%$ of their stellar mass at $z > 2-3$, over a period
of only $\sim 0.1-0.3$ Gyr, when the Universe was only $\sim 4$ Gyr old.
A more recent study of early star formation in these galaxies, however,
raises some doubt about whether or not a single bout of star formation
can fully account for the observed stellar population \cite{25}. In
this paper, we will adopt the conventional view that these galaxies
experienced only minor subsequent episodes of star formation, and
are thus the oldest objects at all redshifts \cite{19}, though our
analysis will need to be updated when these issues are better understood.
So, for example, the stellar population in such a galaxy at $z\sim 1$
(i.e., when the Universe was $\sim 7$ Gyr old) formed during the first
$\sim 2-10\%$ of its evolution. And since one measures only the local
difference in redshift between these galaxies, one avoids the need of
pre-assuming a cosmological model, which constitutes a powerful
discriminant for testing different expansion scenarios. The latest
compilation using this approach includes 30 measurements of $H(z)$
over the redshift range $0\lesssim z\lesssim 2$ \cite{2}.

These measurements are based on the observed 4,000 \AA~break in the
passively evolving galaxy spectra. For old stars, this break is a
discontinuity in the spectral continuum due to metal absorption
lines whose amplitude scales linearly with the stellar age and
metal abundance \cite{2}. When the metallicity of these stars is
known, one can measure the age difference $\Delta t$ of two
adjacent galaxies in proportion to the difference of their 4,000
\AA~amplitudes. The metallicity determines the slope of this relation.
And introducing the redshift difference $\Delta z$ of these galaxies,
one can then determine the Hubble constant using the simple relation
\begin{equation}
H(z) = -{1\over (1+z)}\,{dz\over dt}
\approx-{1\over (1+z)}\,{\Delta z\over \Delta t}\;.
\end{equation}
A caveat with this procedure, however, is that several factors may
limit the accuracy with which the differential age of these systems
is measured. Nonetheless, extensive tests \cite{2} have
demonstrated that the 4,000 \AA~feature is mostly dependent on
the age and metallicity of the host galaxies, relying only weakly
on the star formation history, their initial mass function, along
with possible progenitor biases, and (the so-called) $\alpha$-enhancement.
These early, passive galaxies apparently have higher ratios of $\alpha$
elements to iron than the Milky Way. There may also be a progenitor bias
due to an evolution in the mean redshift of galaxy formation as a function
of redshift.

The tests that have been conducted to this point reveal that only an
uncertainty in the metallicity contributes a systematic error
$\sigma_{\rm sys}$ comparable to the statistical errors in the sample.
The progenitor bias contributes at most only a few percent to
$\sigma_{\rm sys}$, and the initial mass function has no impact.
Using a Chabrier or Salpeter initial mass function for all reasonable
metallicities produces a difference of less than $0.3\%$ between the
4,000 \AA~amplitudes estimated for a single stellar population.
The difference is less than $0.2\%$ for solar metallicity \cite{2}.
Likewise, the $\alpha$-enhancement produces an average difference
in the 4,000 \AA~amplitudes of only $\sim 0.5\%$.

Simulations have also shown that variations in the assumed star forming
rate may produce $\lesssim 13\%$ errors in the measured value of $\Delta z$
from measurements of the 4,000 \AA~amplitudes \cite{2}.
All these effects together contribute an overall error of about
$20\%$ to $\Delta z$, and hence the inferred value of $H(z)$. The 30
measurements of $H(z)$ based on the cosmic chronometer approach are
shown in figure~1 \cite{2,6,20,21,22,23}. Also shown in this figure
is the function $H(z)$ reconstructed from the cosmic chronometer data
using Gaussian Processes, along with the $1\sigma$ and $2\sigma$
confidence regions, which we shall discuss shortly.

Before introducing the models, however, we address an issue with
the data illustrated in figure~1 that has raised some concern
recently regarding whether or not the associated errors are being
over-estimated (see, e.g., refs.~\cite{12,25}). The dispersions
shown here are calculated in quadrature from the statistical
($\sigma_{\rm stat}$) and systematic ($\sigma_{\rm sys}$) errors,
the latter of which are assumed to be uncorrelated. But this
assumption may not be valid, given that some contributions to
$\sigma_{\rm sys}$, e.g., the metallicity variations with redshift,
are not truly random \cite{25}. It is safe to say that, at best,
$\sigma_{\rm sys}$ may have both correlated and uncorrelated
components, in which case, the errors shown in figure~1 are
too large. One can see this directly by comparing the published
errors with the deviations of the data relative to the reconstructed
$H(z)$ function, and from the fact that the reduced $\chi^2$ is 
notably smaller than one ($\chi^2_{\rm dof}\sim 0.52$).

To examine the impact of this error over-estimation on model 
selection, in this paper we will therefore also consider a 
modified set of data extracted from the observations shown in
figure~1 but with reduced dispersions. The ideal way to
reduce the errors would be to deduce the fraction $f_s$
representing the degree of correlation in the systematic
errors with which one would then estimate the `true' error as 
\begin{equation}
\sigma(z_i)=\sqrt{\sigma_{\rm stat}(z_i)^2+f_s\,\sigma_{\rm sys}(z_i)^2}\;.
\end{equation}
Presumably, $f_s=0$ corresponds to the systematic errors being 
fully correlated, and $f_s=1$ totally random. As of now, however, 
there are several limitations that prevent us from using this approach.
The first is that we do not know $f_s$ a priori, nor whether this
is even independent of redshift. Second, of the 30 data points
shown in figure~1, only 17 have published values of $\sigma_{\rm stat}$
and $\sigma_{\rm sys}$. We have attempted to reconstruct the
$H(z)$ function using only this reduced set of measurements, but
the Gaussian-process approach is unstable due to the sparseness of
the data, producing unphysical oscillations with redshift. For now, 
we shall instead follow the simpler suggestion described in ref.~\cite{12},
based on the idea that if correctly estimated, a reasonable model
for the data in figure~1 (say, $\Lambda$CDM) should fit these
measurements with a reduced $\chi^2_{\rm dof}$ approximately equal
to one. Wei et al. \cite{12} found that to attain this goal, the
errors shown in figure~1 must be reduced, on average, by about $25\%$.
The $H(z)$ data with their modified error bars are shown in figure~2,
together with the reconstructed $H(z)$ function based on these reduced
dispersions.

\begin{figure}
\begin{center}
\includegraphics[width=5.3in]{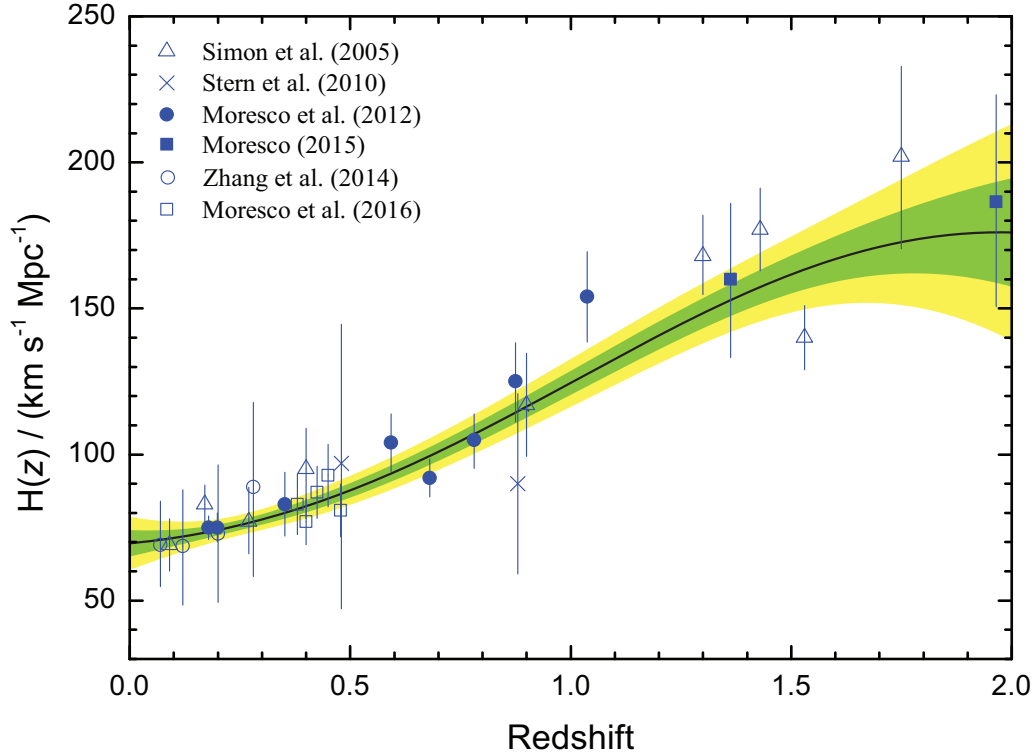}
\end{center}
\caption{Same as figure~1, except that the error bars have here been
reduced by an average of $25\%$. This modification allows a reasonable
cosmological model, such as $\Lambda$CDM, to fit the data with a reduced
$\chi^2_{\rm dof}\approx 1$ \cite{12}. Also shown here is the function
$H(z)$ reconstructed with Gaussian Processes (solid black). The green
and yellow shaded regions represent the $1\sigma$ and $2\sigma$ confidence
regions, respectively, of the reconstruction.}
\label{figure2}
\end{figure}

\subsection{Gaussian Processes}
The principal benefit of using the Gaussian Processes (GP) approach \cite{5} is
that it avoids having to assume a parametric form of the function representing
the data based on particular models that may, or may not, be reasonable representations
of the true redshift-dependence of the measurements. A complete description of this
method has been provided in ref.~\cite{5}, and a catalog of related algorithms may
be downloaded from a website maintained by these 
authors.\footnote{http://www.acgc.uct.ac.za/$\sim$seikel/GAPP/index.html} 

Gaussian Processes can model a function f(x) rigorously without assuming 
any prior parametric form. One assumes that the $n$ observations of a data set
y=$\{y_1,y_2,....,y_n\}$ were sampled from a multivariate Gaussian distribution,
which then allows these data to be partnered with Gaussian Processes (GP). Most 
often, the mean of this partner GP is assumed to be zero. But though modeling the 
data using Gaussian Processes is straightforward, there are nonetheless two 
possible areas of ambiguity with this technique, so we 
highlight these here and describe steps we have taken to ensure that the outcome
of our reconstruction is not heavily biased by our choice of GP components. One
of these has to do with the fact that the values of the function evaluated at 
different points $x_1$ and $x_2$ are not independent of each other. As such, 
one introduces a covariance function $k(x_1,x_2)$ to handle the linkage between
neighboring points, though the form of $k(x_1,x_2)$ is not unique or well
known. In principle, there is a broad range of such covariance functions, 
and while it makes sense to pick one that depends only on the distance between 
different data points, this is actually not required. It is common in this type 
of work to select a squared exponential function for this purpose, 
\begin{equation}
k(x_1,x_2) = \sigma_f^2\exp\left(-{(x_1-x_2)^2\over 2l^2}\right)\;,
\end{equation}
simply because it is infinitely differentiable and useful for reconstructing 
also the derivative of a function that represents the data. In contrast to
actual parameters, the so-called hyperparameters $\sigma_f$ and $l$ do not
specify the form of the function, only its ``bumpiness." The characteristic
length $l$ represents a distance in $x$ over which the reconstructed function
varies significantly, while the signal variance $\sigma_f$ scales this
dependence in the ordinate direction. The covariance matrix obtained 
using Equation~(2.3) for $\{x_1,x_2,....x_n\}$ observation points is
\begin{equation}
K=
\begin{bmatrix}
k(x_1,x_1) & k(x_1,x_2) & ..... & k(x_1,x_n)\\
k(x_2,x_1) & k(x_2,x_2) & ..... & k(x_2,x_n)\\
.......... & .......... &...... & ..........\\
k(x_n,x_1) & k(x_n,x_2) & ..... & k(x_n,x_n)
\end{bmatrix}\;.
\end{equation}
For a new observation point $x_*$, one also needs the vector
\begin{equation}
K_*\equiv
\begin{bmatrix}
k(x_*,x_1) & k(x_*,x_2) & ..... & k(x_*,x_n)
\end{bmatrix}\;,
\end{equation}
and the point $K_{**}\equiv k(x_*,x_*)$. As we have pointed out, the 
data may be represented as a sample from a multivariate GP, such that
\begin{equation}
\begin{bmatrix}
y \\ y_* 
\end{bmatrix}
= N\bigg(0,
\begin{bmatrix}
 K & K_*^T \\
 K_*   & K_{**}
\end{bmatrix}
\bigg)\;.
\end{equation}
In obtaining this, one must maximize the conditional probability
\begin{equation}
p(y_*|y)\sim N(K_*K^{-1}y,K_{**}-K_*K^{-1}K_*^T)\;.
\end{equation}
The mean of the distribution is an estimate of $y_*$, which is given as
\begin{equation}
\mu(y_*)=K_*K^{-1}y\;,
\end{equation}
and the uncertainty of the estimate is given as
\begin{equation}
 {\rm var}(y_*)=K_{**}-K_*K^{-1}K_*^T\;.
\end{equation}

All of the results discussed in this paper, particularly those shown in 
figs.~1-6, are based on the use of the kernel given in Equation~(2.3). 
However, to ensure that this choice of covariance function is not 
unduly affecting our model selection, we also carry out a parallel
set of simulations in the Appendix based on the use of a very different
kind of kernel known as a Mat\'ern covariance function, specifically
the one called Mat\'ern92 \cite{5}. As described in the Appendix, the 
choice of kernel may change the $p$-values by a few points, but the outcome
of model selection based on this approach is not changed qualitatively.
The rank ordering of models we consider here (see below) appears to be
unaffected by the choice of covariance function. 

A second possible area of ambiguity has to do with the values of the
hyperparameters themselves. The common approach followed in this context
is to train them by maximizing the likelihood that the reconstructed function
has the measured values at the data points $x_i$. Of course, for a purely
Bayesian analysis, one should marginalize over the hyperparameters instead
of optimizing them, but for most applications (as we have here), the
marginal likelihood is sharply peaked. Optimization is therefore a good
approximation to marginalization, so for the purpose of the model selection
we describe in this paper, there is no freedom to choose $\sigma_f$ and $l$
separately. 

\subsection{Reconstructed $H(z)$ Function versus the Data}
Let us now pause briefly to discuss a very salient point emerging from the
reconstructed $H(z)$ functions shown in figures~1 and 2. As alluded to above,
cosmic chronometer measurements of the Hubble constant have themselves been
used in recent years, primarily to optimize the parameters in the standard
model $\Lambda$CDM, but also on occasion to compare the predictions
of $\Lambda$CDM with those of the $R_{\rm h}=ct$ universe \cite{7,8}.
In the latter, the power-law form of the expansion rate, i.e.,
$H(z)=H_0(1+z)$, has been shown to fit these data better than the
variable rate in the standard model. But those results were the outcome
of model fitting to the data. Figure~1 provides us with a new perspective
on this comparison, because the Hubble constant reconstructed from the data
using Gaussian Processes is entirely free of any presumed model or assumed
fitting function. And this approach clearly demonstrates that the $(1+z)$
power law is a much better representation of the reconstructed $H(z)$ function
than the variable rate predicted by the standard model (see fig.~3
below). This outcome should not be underestimated, particularly in view of
other recent attempts at showing that the cosmic chronometer data support
the inference that the Universe underwent a transition from deceleration
to acceleration at a redshift $z\sim 0.5-0.7$ \cite{2}. These arguments
are based on the adoption of specific empirical functions to fit the data,
unlike the Gaussian Process reconstruction which makes no such assumptions.
The reconstructed $H(z)$ in figure~1 shows no evidence of such a transition,
validating the conclusions drawn earlier in refs.~\cite{7,8}. As we shall
see shortly, however, the function $H(z)$ reconstructed from the data with reduced
dispersions (fig.~2) is not as compelling as figure~1 in this regard, principally
because none of the models fits it as well as in the first case. Nonetheless,
the model rankings are not changed, so the two reconstructions agree qualitatively,
if not quantitatively.

\section{Model Comparisons}
In this section, we will now briefly introduce the 6 different
cosmological models we intend to test against the reconstructed
$H(z)$ functions, including the concordance model (based on the
{\it Planck} optimized parameters) and the $R_{\rm h}=ct$ universe.
The models we will compare are the following:

\begin{enumerate}
\item The $R_{\rm h}=ct$ universe (a Friedmann-Robertson-Walker cosmology
with zero active mass; \cite{16,17}). The basis for this model is the
total equation of state $\rho+3p=0$, where $\rho$ and $p$ are the total
energy density and pressure of the cosmic fluid \cite{9,11,16,17}. This
cosmology has only one free parameter---the Hubble constant $H_0$, and
\begin{equation}
H^{R_{\rm h}=ct}(z) = H_0(1+z)\;.
\end{equation}

\item The concordance model, based on the {\it Planck} optimization of
the parameters in $\Lambda$CDM. This model has the Hubble function
\begin{equation}
H^{\Lambda{\rm CDM}}(z)= H_0\left[\Omega_{\rm m}(1+z)^3+\Omega_{\rm r}(1+z)^4+
\Omega_\Lambda\right]^{1/2}\;,
\end{equation}
where $\Omega_i$ is the energy density $\rho_i$ of species $i$, for
radiation ($\Omega_{\rm r}$), matter ($\Omega_{\rm m}$) and dark energy
($\Omega_{\Lambda}$), scaled to the critical density, $\rho_c\equiv 3c^2
H_0^2/8\pi G$. Its parameters have the (fixed) prior values $H_0=67.4\pm1.4$ 
km s$^{-1}$ Mpc$^{-1}$, $\Omega_{\rm m}=0.314\pm0.020$, and $\Omega_\Lambda=
0.686\pm0.020$ \cite{26}. Note that $\Omega_{\rm r}\ll 1$ in the redshift 
range associated with cosmic chronometers, and $\Omega_{\rm m}+\Omega_{\Lambda}$ 
(with $\Omega_{\rm r}=0$) was fixed to $1$ because this parameter optimization 
represents a flat universe. We therefore do not include a spatial curvature term 
$\Omega_{\rm k}$ in Equation~(3.2).

\item The flat $\Lambda$CDM ($\Omega_{\rm m}$) model, with matter and dark-energy
densities fixed by the condition $\Omega_\Lambda=1-\Omega_{\rm m}$
(when radiation is insignificant). The predictions of this model are
based on Equation~(3.2) with the optimization of one free parameter,
$\Omega_{\rm m}$.

\item The flat $\Lambda$CDM ($w$) model, with matter and dark-energy densities
fixed by the condition $\Omega_{\rm de}=1-\Omega_{\rm m}$ (again,
when radiation is insignificant). Here, $H_0$ and $\Omega_{\rm m}$ have
prior values (from {\it Planck}), but the dark-energy equation of
state, $w_{\rm de}\equiv p_{\rm de}/\rho_{\rm de}$, where $p_{\rm de}$
is the dark-energy pressure, is unconstrained. Also,
\begin{equation}
H^{w{\rm CDM}}(z)= H_0\left[\Omega_{\rm m}(1+z)^3+\Omega_{\rm r}(1+z)^4+
\Omega_{\rm de}(1+z)^{3(1+w_{\rm de})}\right]^{1/2}\;.
\end{equation}

\item The flat $\Lambda$CDM ($H_0$) model, with matter and dark-energy
densities fixed by the condition $\Omega_\Lambda=1-\Omega_{\rm m}$
(when radiation is insignificant). Its predictions are based on Equation~(3.2)
with an unconstrained $H_0$.

\item Einstein--de Sitter space, which contains only matter. This model
has only one free parameter, $H_0$, and
\begin{equation}
H^{\rm EdS}(z)=H_0(1+z)^{3/2} \;.
\end{equation}

\end{enumerate}

\begin{figure}[H]
\begin{center}
\begin{tabular}{cc}
\hskip-0.4in\includegraphics[width=0.62\linewidth]{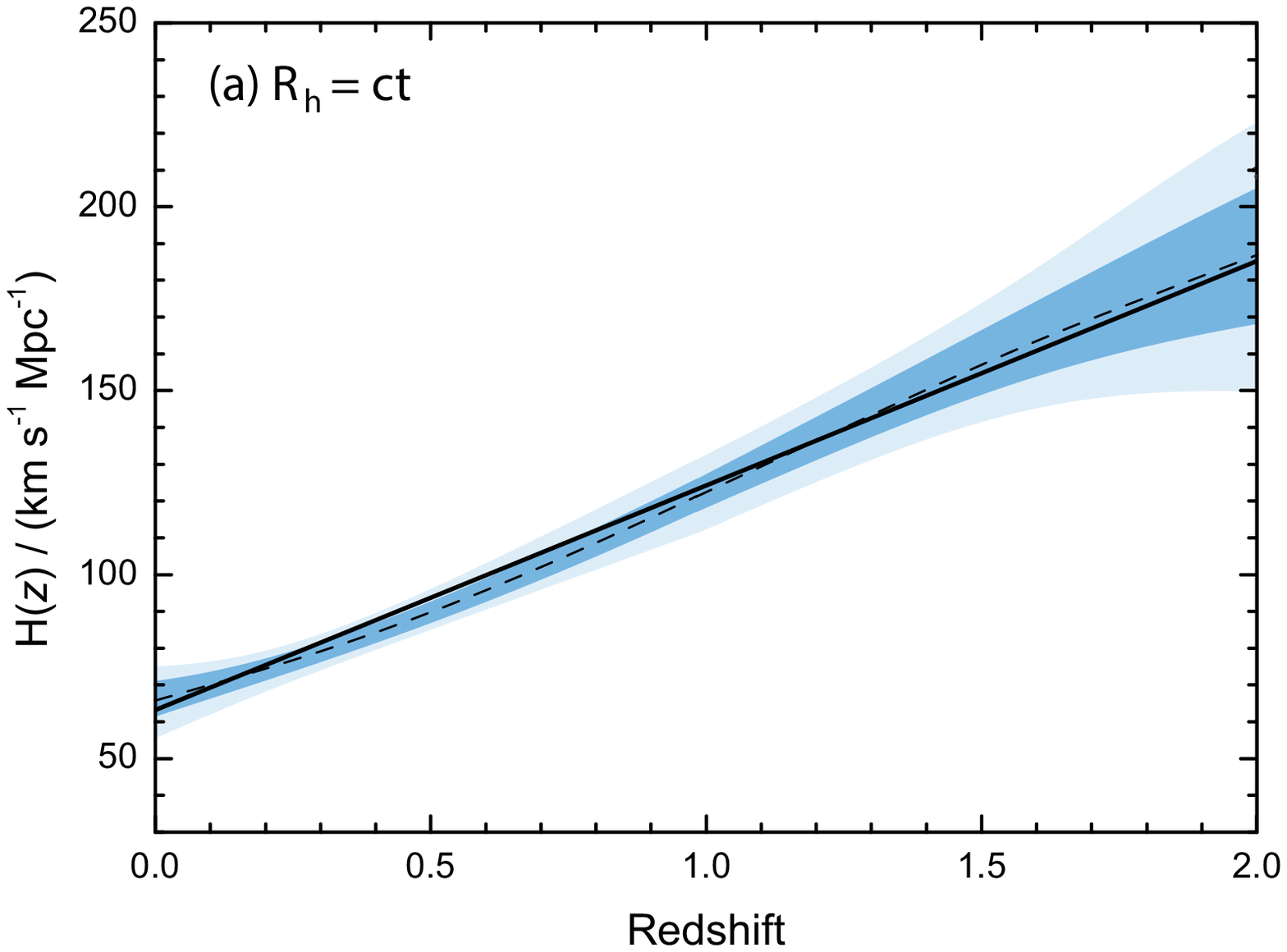}
\hskip-0.6in\includegraphics[width=0.62\linewidth]{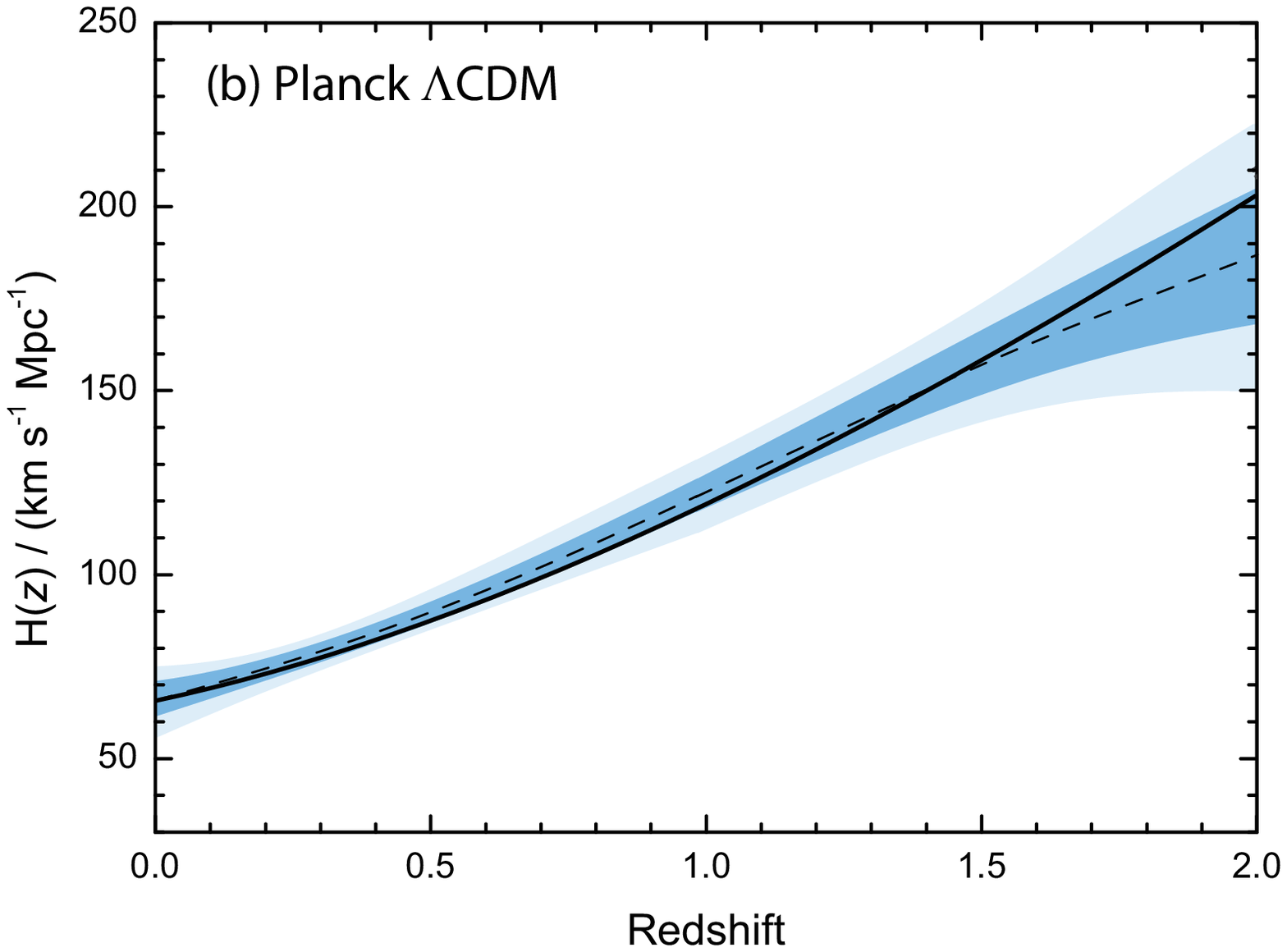} \\
\hskip-0.4in\includegraphics[width=0.62\linewidth]{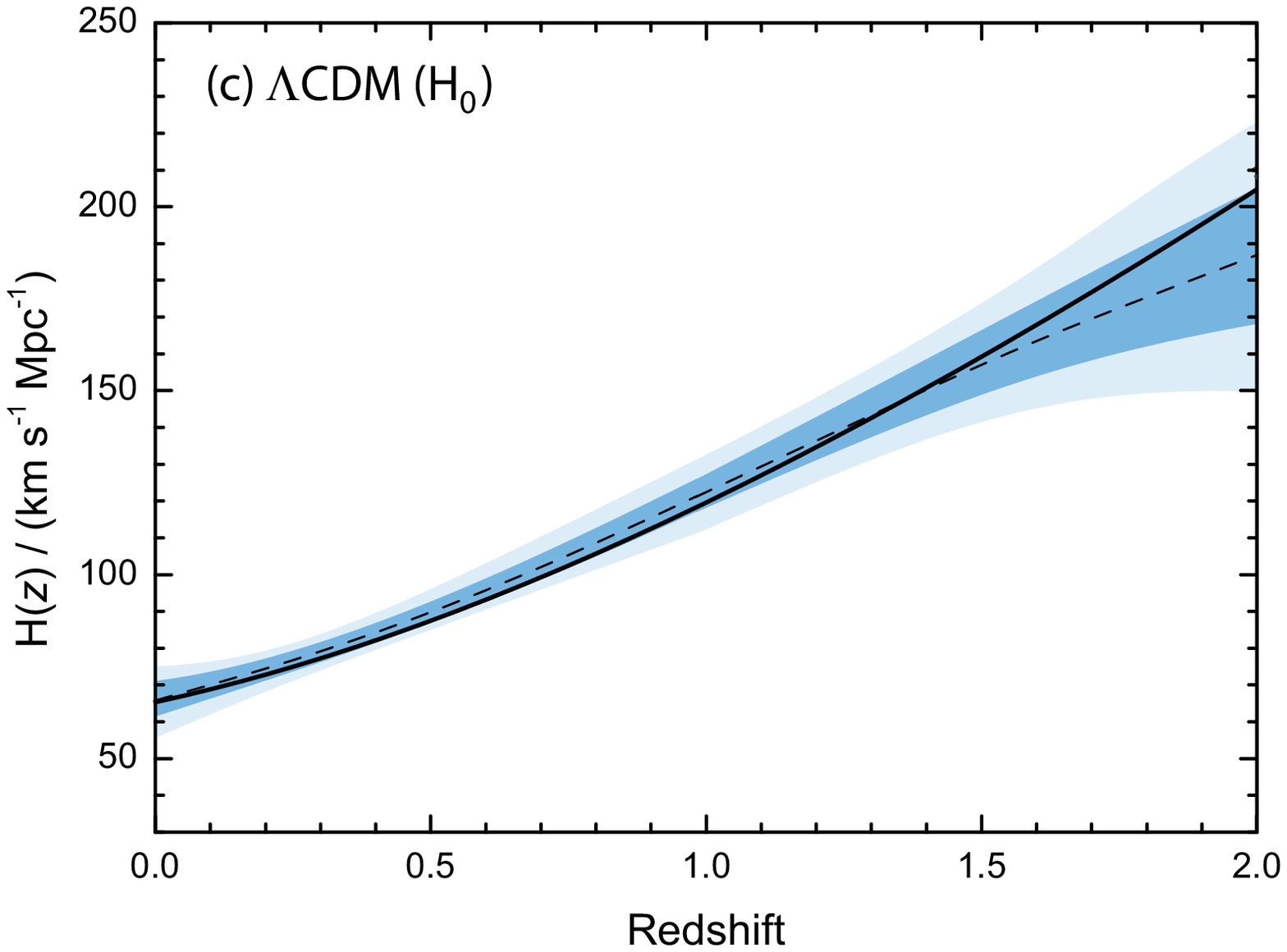}
\hskip-0.6in\includegraphics[width=0.62\linewidth]{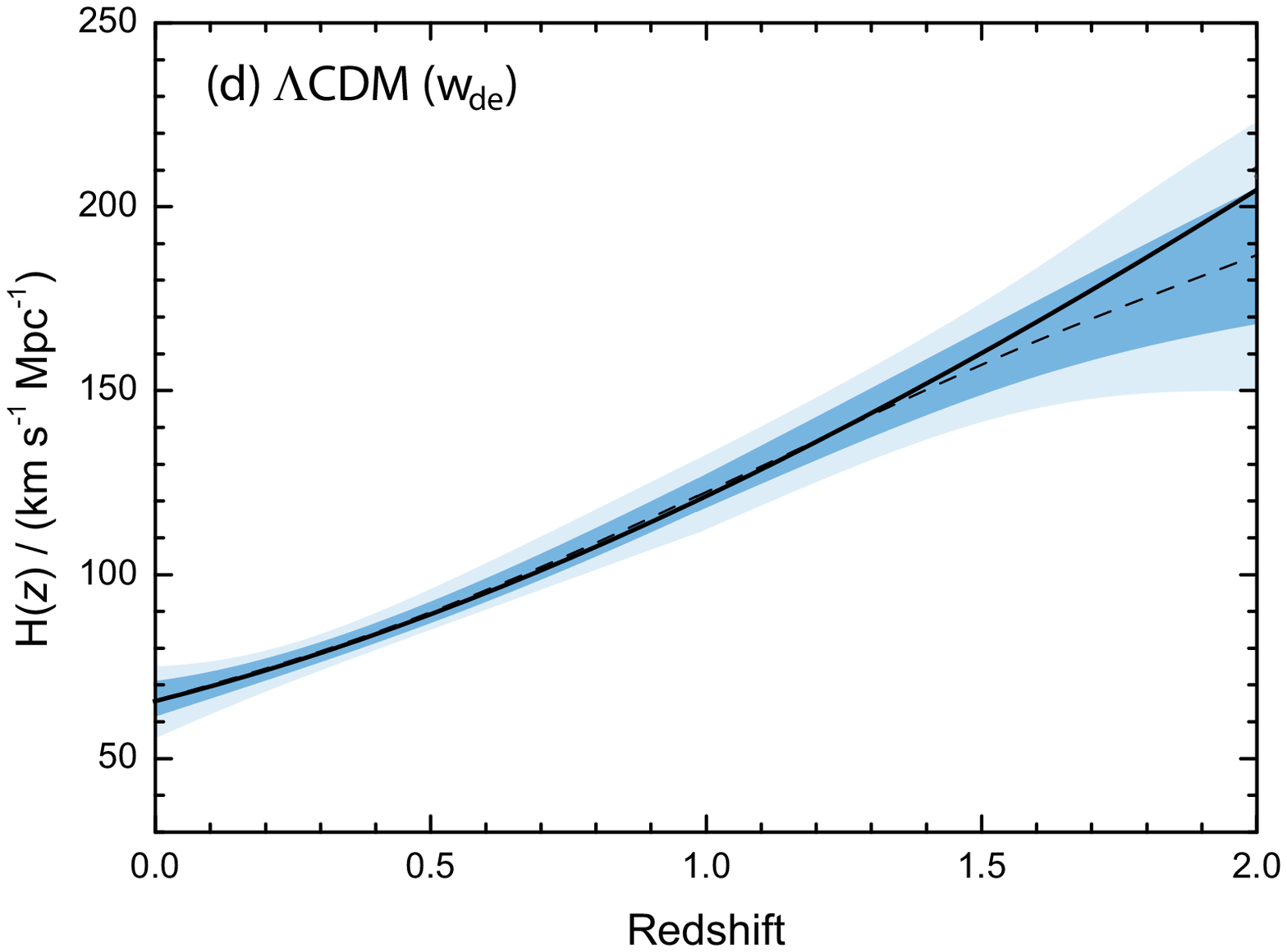} \\
\hskip-0.4in\includegraphics[width=0.62\linewidth]{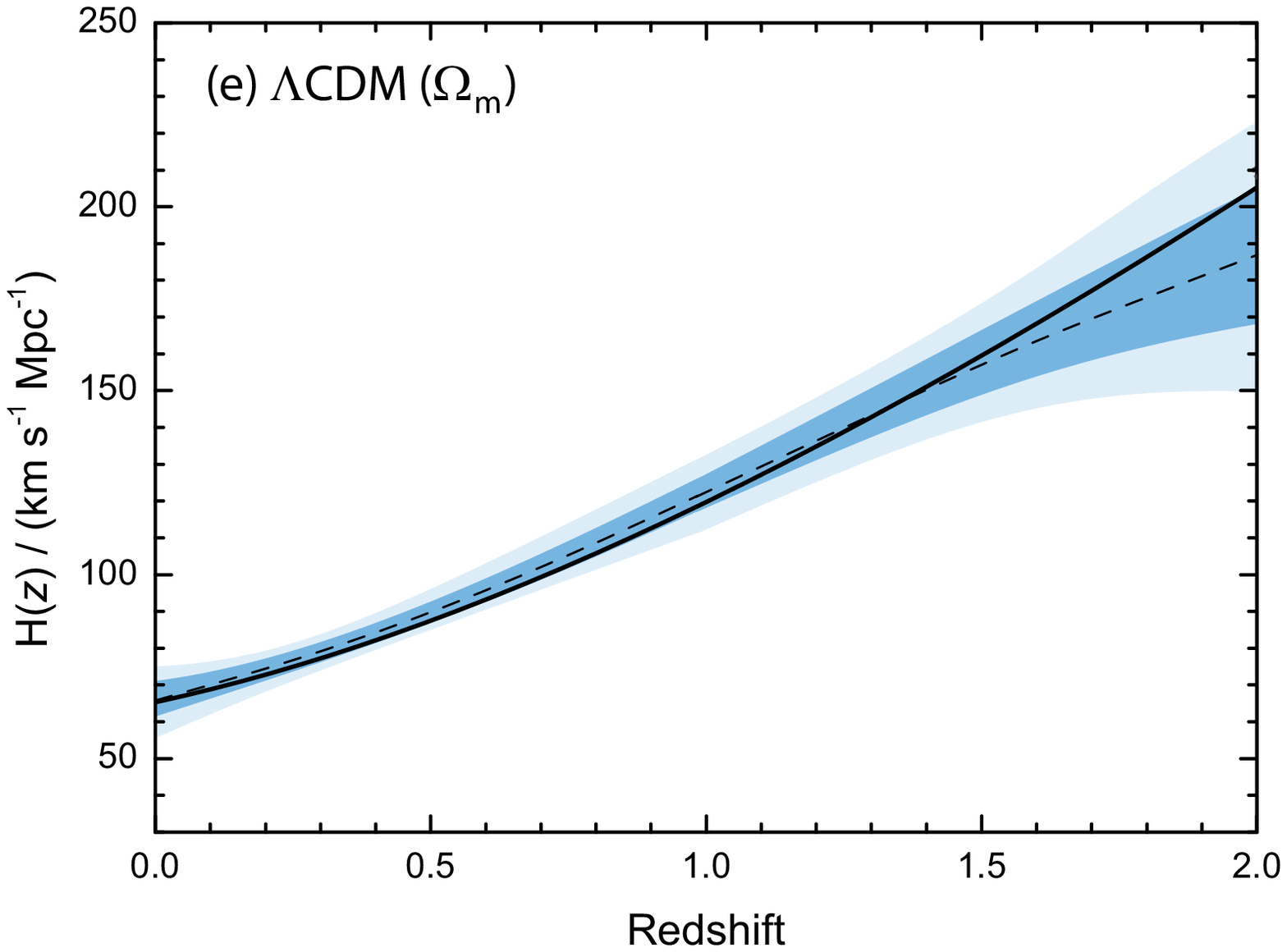}
\hskip-0.6in\includegraphics[width=0.62\linewidth]{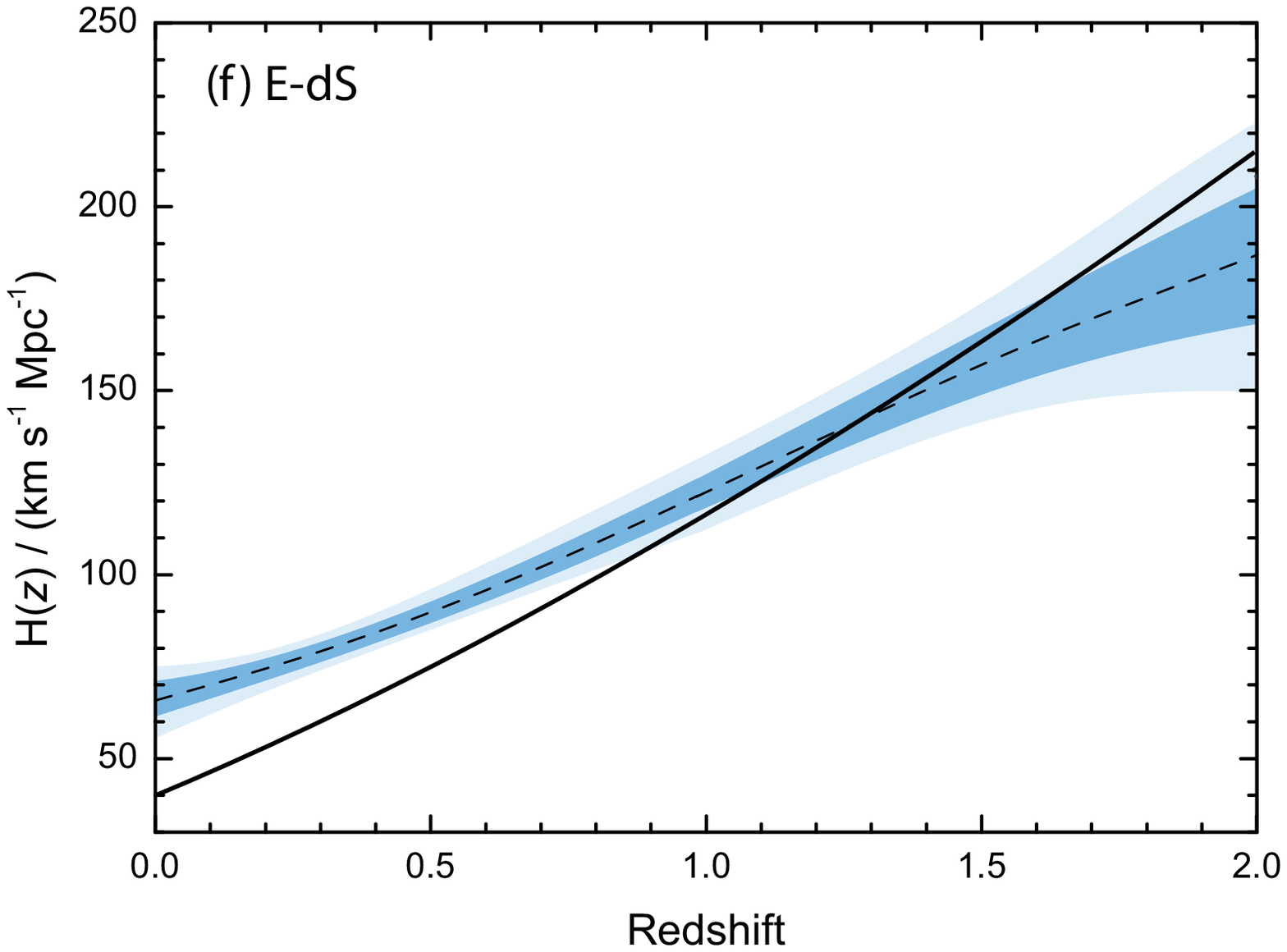}
\end{tabular}
\end{center}
\caption{The Hubble constant $H(z)$ (solid curve) in various cosmologies
optimized to fit the reconstructed function in figure~1 (dashed curve): 
(a) The $R_{\rm h}=ct$ universe; (b) Planck $\Lambda$CDM; (c) $\Lambda$CDM ($H_0$), 
with a re-optimized (i.e., re-fitted) Hubble parameter; (d) flat
$\Lambda$CDM ($w_{\rm de}$); (e) $\Lambda$CDM ($\Omega_{\rm m}$);
(f) Einstein de Sitter. The blue bands indicate the $1\sigma$ (dark 
shade) and $2\sigma$ (light shade) confidence regions from figure~1.}
\label{figure3}
\end{figure}

Starting with the $H(z)$ function reconstructed using the full sample
of 30 measurements (fig.~1), one may gauge how well it compares to model predictions
by examining the theoretical best-fit curves (more on this below)
in relation to the $1\sigma$ and $2\sigma$ confidence regions shown in figure~3.
For the sake of clarity, the reconstructed $H(z)$ curve itself is shown as a thin
dashed line in these panels. A quick inspection by eye suggests that the
$R_{\rm h}=ct$ curve is very close to the reconstructed function, while the other
models all have some excess curvature, either at low or high redshifts, introducing
at least some tension with the observations.

To quantify these comparisons, we will adopt the following procedure.  We use
the data and their $1\sigma$ errors plotted in figure~1 to create mock samples
of $30$ values of the Hubble constant with the same redshifts, $z_i$
($i=1,...,30$), as the measurements, but with Gaussian randomized
values $H^{\rm mock}(z_i)=H(z_i)+r\sigma_i$, where $r$ is a
Gaussian random variable with mean $0$, and variance $1$, and $\sigma_i$
is the dispersion at $ z_i$ (see figure~1).

\begin{figure}
\begin{center}
\includegraphics[width=4.3in]{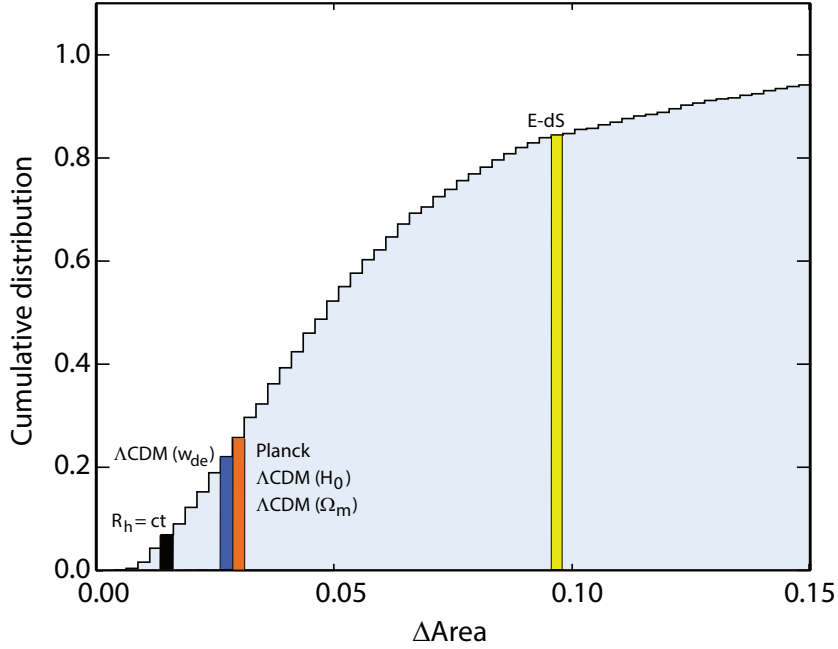}
\end{center}
\caption{Cumulative probability distribution (normalized to $1$) of the differential
area calculated for $H(z)$ according to Equation~(3.5) for mock samples
constructed via Gaussian randomization of the measured $H(z_i)$ values (see
text preceding Equation~3.5). The various models are shown according to their 
probabilities listed in Table~1.}
\label{figure4}
\end{figure}

Using these mock data, we reconstruct a mock $H^{\rm mock}(z)$
function, and then calculate a normalized absolute area difference
between this and the real function, according to
\begin{equation}
\Delta A=\int_0^2 dz\,\left|H^{\rm mock}(z)-H(z)\right|\bigg/
\int_0^2 dz\,H(z)\;.
\end{equation}
With this approach, we build a distribution of frequency versus differential area
\cite{27}, from which we then construct the cumulative probability distribution shown in
figure~4.

For each model $i$, we calculate the differential areas analogous to Equation~(3.5),
replacing $H^{\rm mock}$ with its model specific function $H^i(z)$, and then estimate
the probability (i.e., the $p$-value) of this cosmology being consistent with the
reconstructed $H(z)$ function using figure~4. For a given $\Delta A$, this 
figure shows what fraction of the randomized realizations had a differential area 
smaller than this value. Thus, a model's calculated $\Delta A$ shows the probability 
(the $p$-value) that this cosmology's prediction differs from the reconstructed $H(z)$ 
function due solely to cosmic variance. If a model has a free parameter, 
we optimize its value within its physically meaningful range by finding a minimum 
of the differential area $\Delta A$. For $\Omega_{\rm m}$, this range
is $(0,1)$, while for $w_{\rm de}$ we assume $(-1,0)$, i.e., we exclude phantom
dark energy with $w_{\rm de}<-1$. The resulting probabilities and parameter values 
are quoted in Table~1.

As we expected from a quick inspection of figure~3, the reconstructed 
Hubble constant prefers the $R_{\rm h}=ct$ cosmology, and provides 
reasonable fits for $\Lambda$CDM with either of the parameters
$w_{\rm de}$, $H_0$ or $\Omega_{\rm m}$ re-optimized to fit the
reconstructed $H(z)$ function, but with smaller probabilities.
Quite coincidentally, the optimized value of $\Omega_{\rm m}$
when it is the sole parameter allowed to vary (row 5) is identical to 
the {\it Planck} value (row 4). As such, the probabilities for
Planck $\Lambda$CDM and $\Lambda$CDM ($\Omega_{\rm m}$) are the same.
The Einstein-de Sitter model, however, is strongly disfavoured by
this comparative test.

We repeat this procedure using the data with reduced dispersions (fig.~2)
in order to gauge the impact of correlation in the systematic errors on
the reconstruction of $H(z)$. The individual model fits are shown in
figure~5, with the corresponding cumulative probability distribution
in figure~6. The model rankings and their probabilities are listed in 
Table~2. We notice in this case that none of the models fit the
reconstructed $H(z)$ function very well. Even $R_{\rm h}=ct$, which
is marginally favoured over the others, sits at roughly the halfway
point of all possible Gaussian randomizations based on the reduced
measurement errors, i.e., a simple variance of the reconstructed function 
$H(z)$ would produce a differential area $\Delta A$ smaller than that 
corresponding to $R_{\rm h}=ct$ about half of the time. Though the two 
model rankings (in Tables~1 and 2) are consistent with each other, one 
can see from figure~2 why the second reconstructed $H(z)$ is not fully
consistent with the first. The scatter in the measurements increases with 
redshift, suggesting that a simple uniform reduction (by $25\%$) of the 
errors to account for partial correlation in the systematics may not be
a reasonable approach. But attempting a more detailed mitigation of
the systematic errors is not feasible, given how little is known
at this stage about $\sigma_{\rm stat}$ and, particularly,
$\sigma_{\rm sys}$.

\begin{table*}
\center
  \centerline{{\bf Table 1.} Model comparison using the $H(z)$ function reconstructed from}  
  \centerline{the 30 measurements with published errors. The value of $\Omega_{\rm m}$ 
              is fixed in}
  \centerline{every case, except for the model $\Lambda$CDM $(\Omega_{\rm m})$}\vskip 0.1in
\footnotesize
  \centering
  \begin{tabular}{lccccc}
&& \\
    \hline
\hline
&& \\
Model & $H_0$ & $\Omega_{\rm m}$ & $\Omega_{\rm de}$&$w_{\rm de}$&Prob. $(\%)$ \\
&(km s$^{-1}$ Mpc$^{-1}$)&&&& (Fig.~4) \\
&& \\
\hline
&& \\
$R_{\rm h}=ct$ & $63.0$& --& --& --& $93.0$ \\
$\Lambda$CDM ($w_{\rm de}$) & $67.4$& $0.314$ {\rm (fixed)}& $1-\Omega_{\rm m}$& $-0.913$& $83.0$ \\
$\Lambda$CDM ($H_0$) &$70.5$& $0.314$ {\rm (fixed)}& $1-\Omega_{\rm m}$& $-1$& $78.3$ \\
Planck $\Lambda$CDM&$67.4$ &$0.314$ {\rm (fixed)} &$1-\Omega_{\rm m}$&$-1$ &$74.2$ \\
$\Lambda$CDM ($\Omega_{\rm m}$)& $67.4$& \qquad $0.314$ {\rm (optimized)}& $1-\Omega_{\rm m}$& $-1$& $74.2$ \\
E-dS &$41.3$& $1.0\quad$ {\rm (fixed)}& --& --& $16.6$ \\
&& \\
\hline\hline
  \end{tabular}
\end{table*}

\begin{figure}[H]
\begin{center}
\begin{tabular}{cc}
\hskip-0.4in\includegraphics[width=0.62\linewidth]{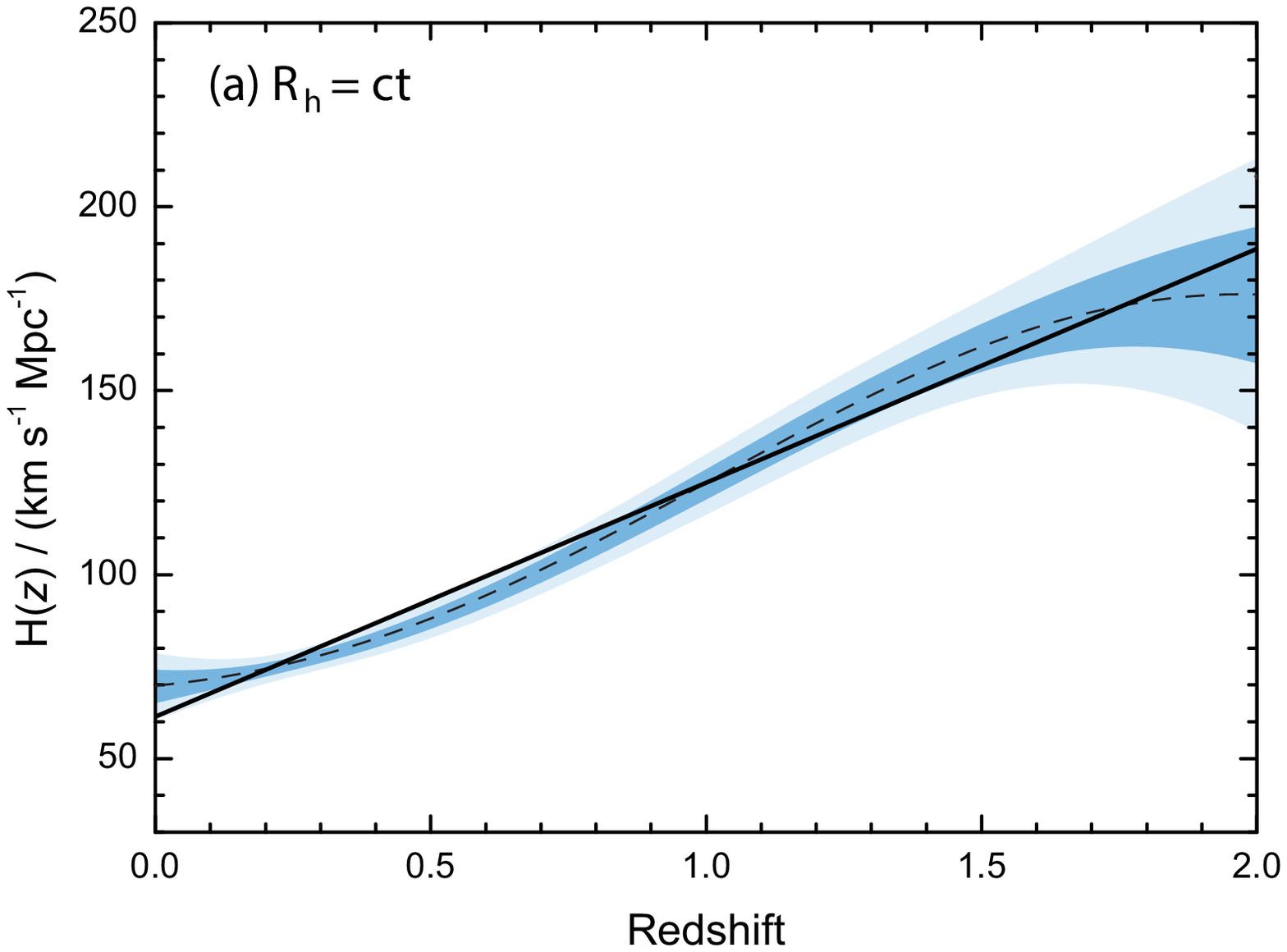}
\hskip-0.6in\includegraphics[width=0.62\linewidth]{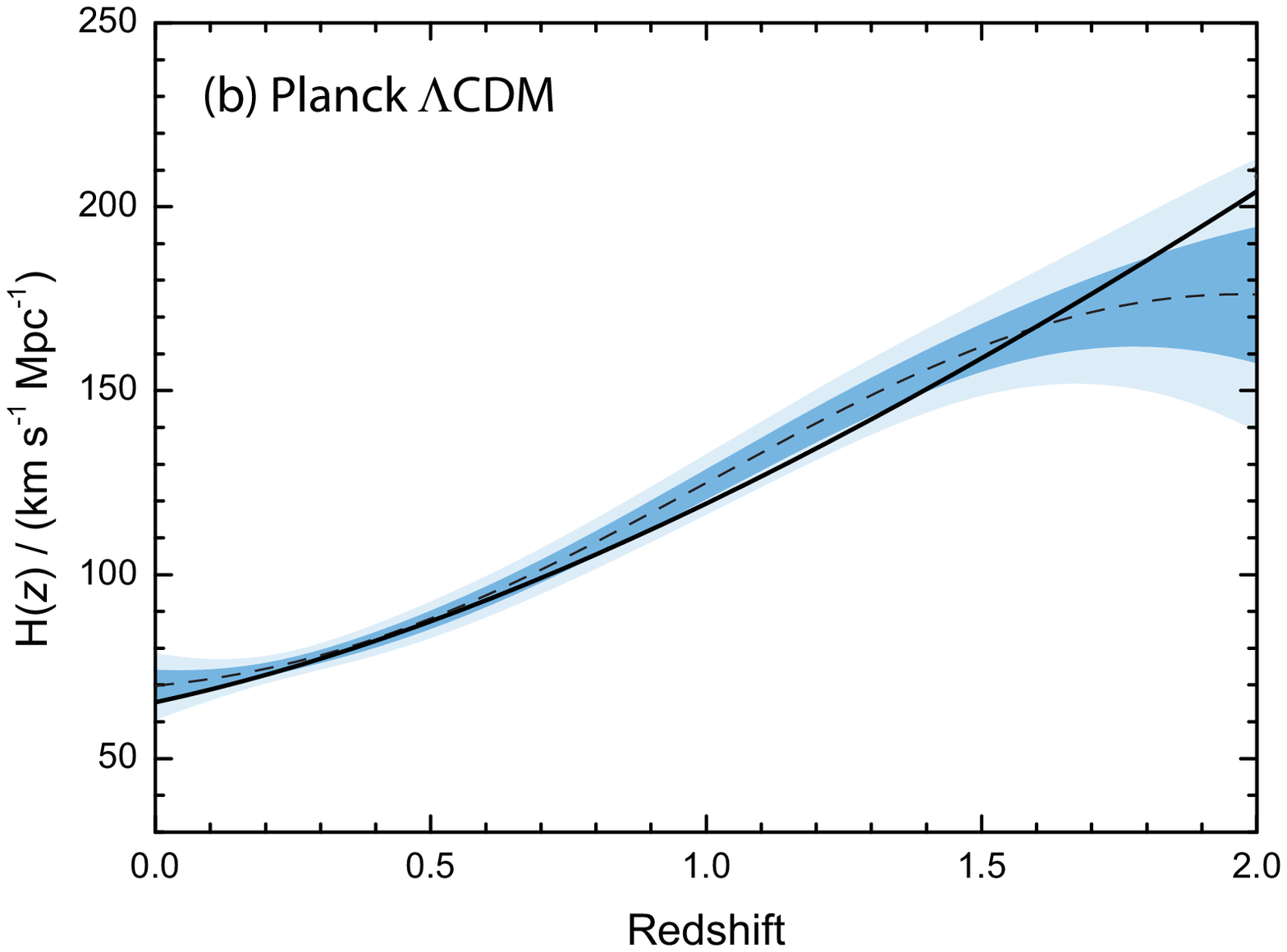} \\
\hskip-0.4in\includegraphics[width=0.62\linewidth]{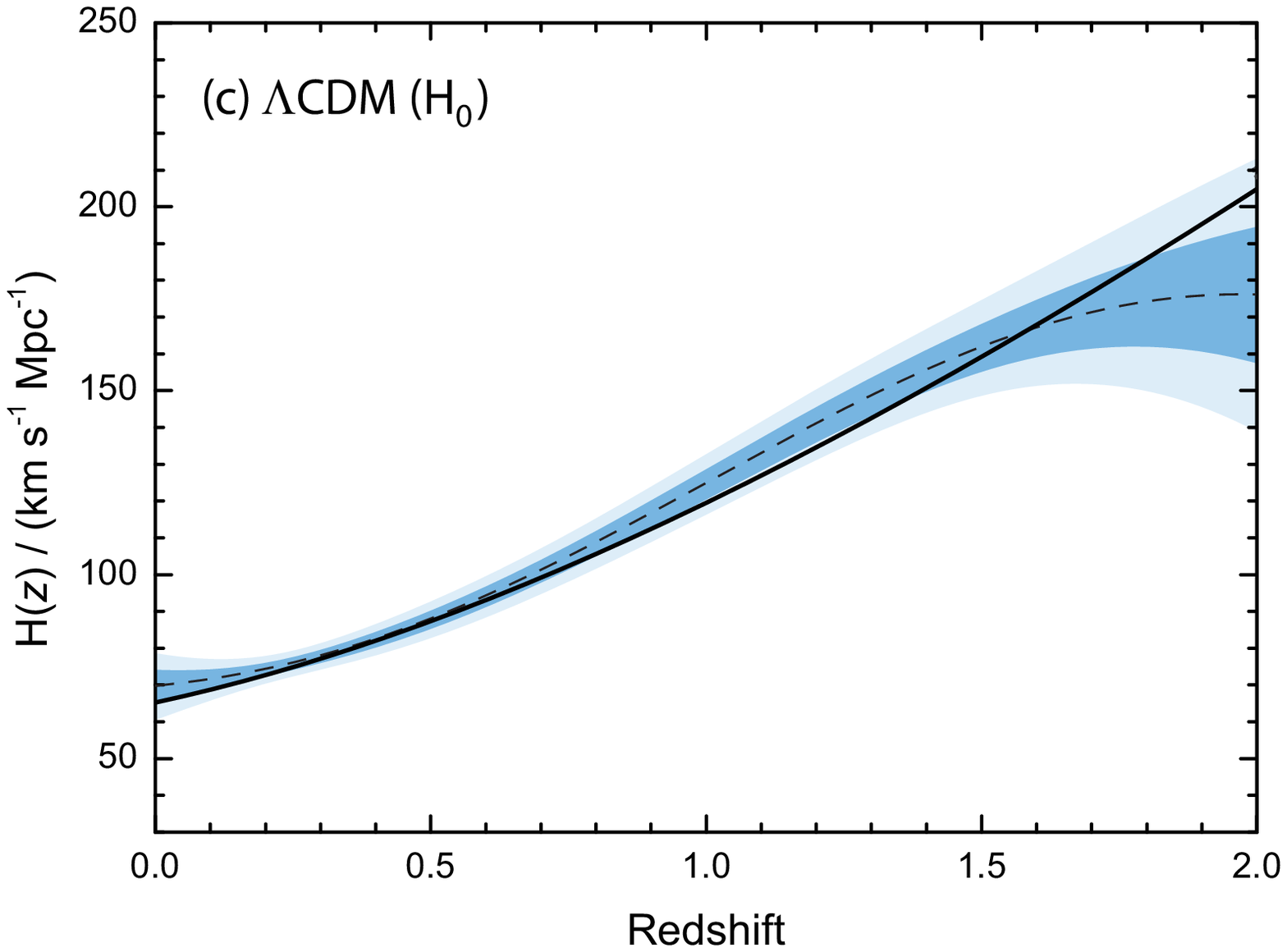}
\hskip-0.6in\includegraphics[width=0.62\linewidth]{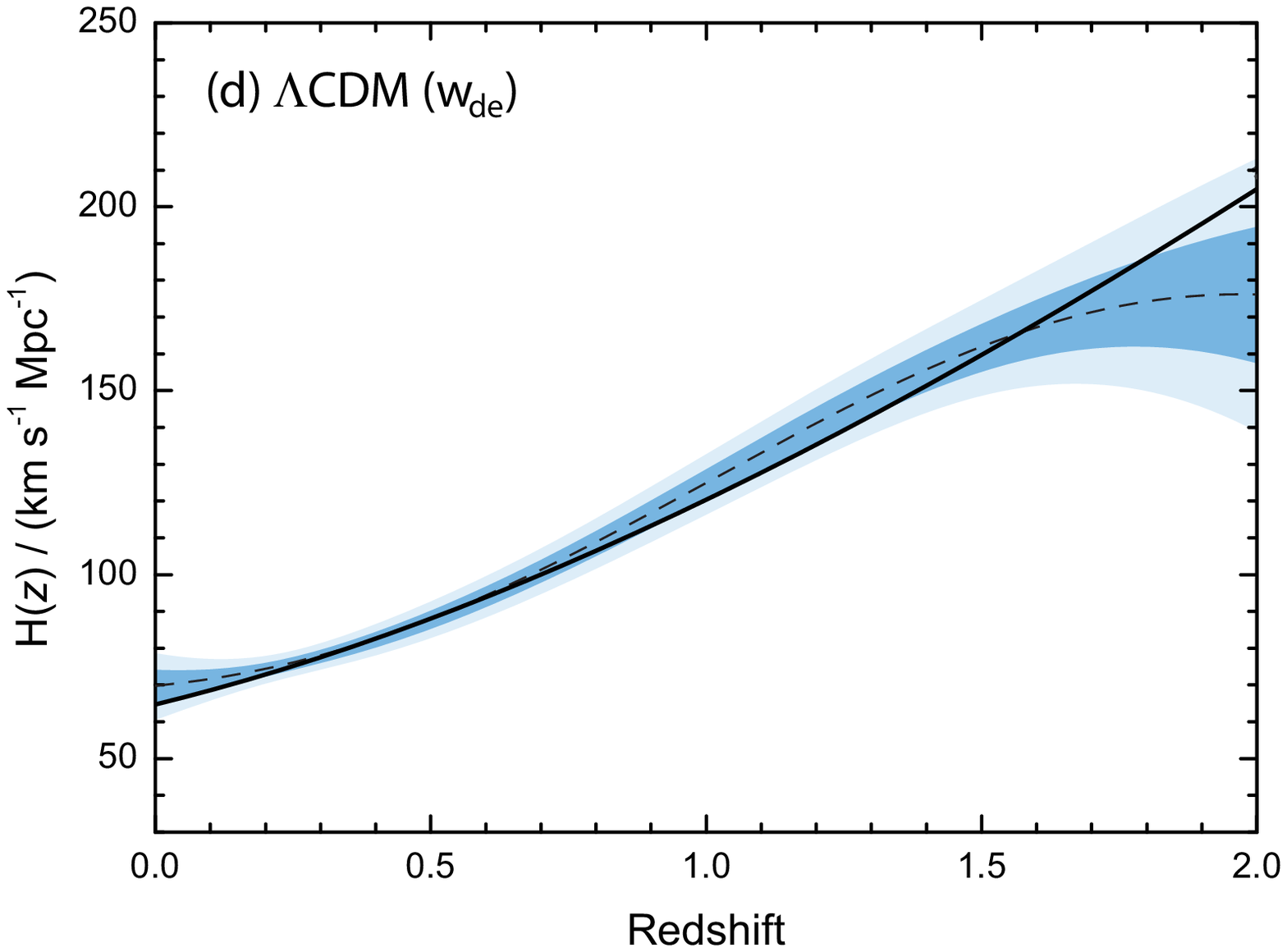} \\
\hskip-0.4in\includegraphics[width=0.62\linewidth]{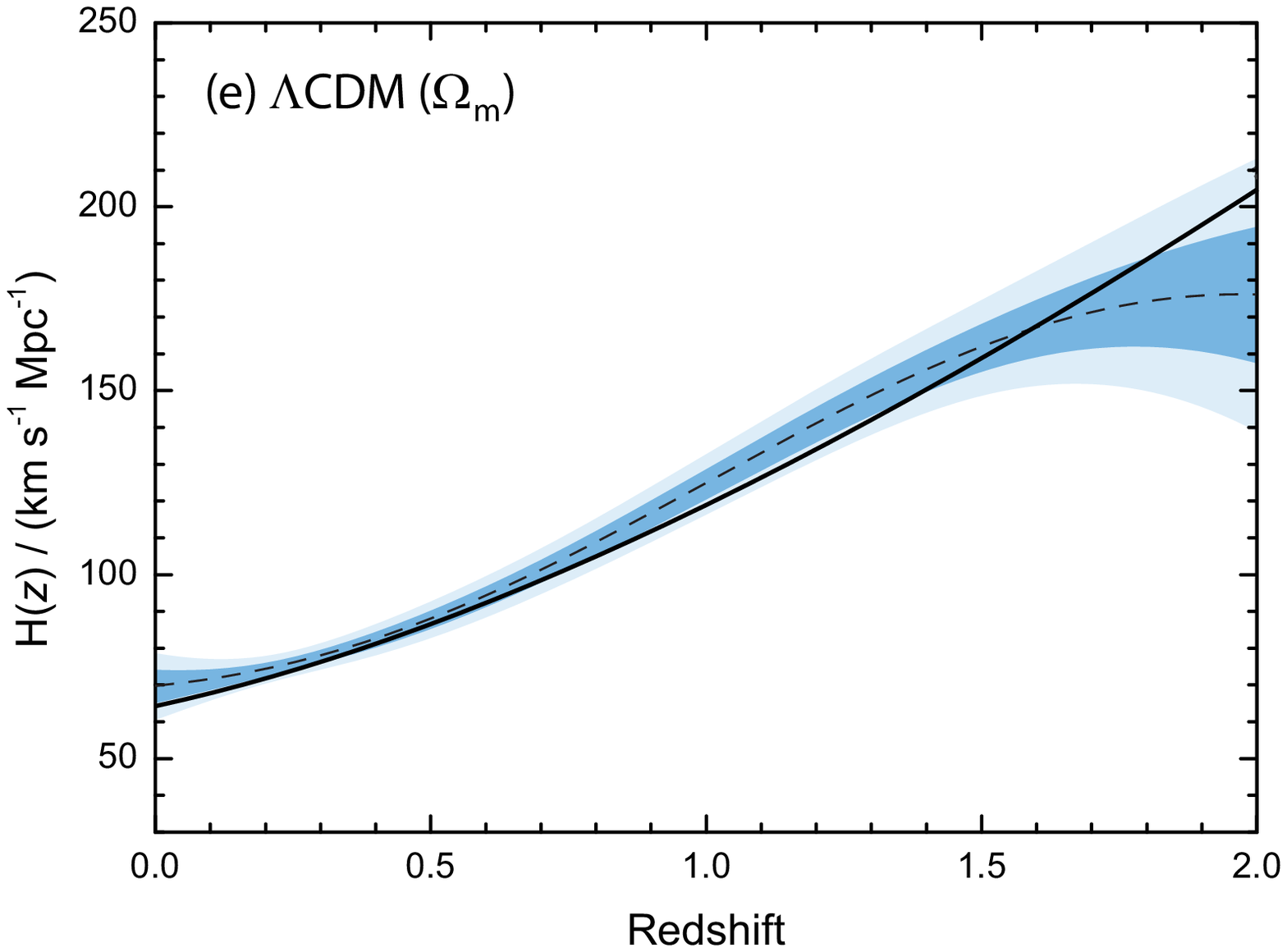}
\hskip-0.6in\includegraphics[width=0.62\linewidth]{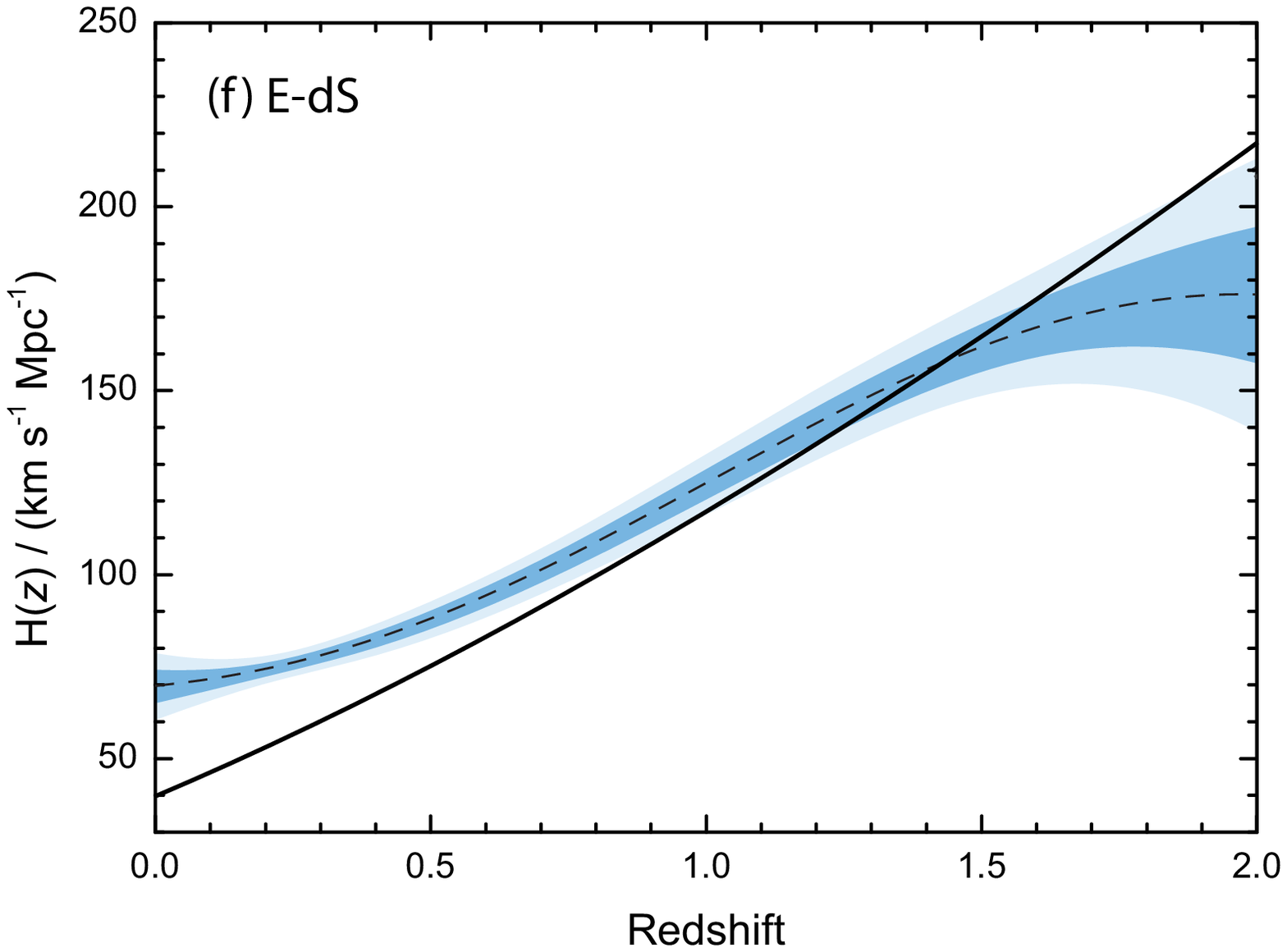}
\end{tabular}
\end{center}
\caption{The Hubble constant $H(z)$ (solid curve) in various cosmologies
optimized to fit the reconstructed function in figure~2 (dashed curve): (a)
The $R_{\rm h}=ct$ universe; (b) Planck $\Lambda$CDM; (c) flat $\Lambda$CDM ($H_0$);
(d) flat $\Lambda$CDM ($w_{\rm de}$); (e) $\Lambda$CDM ($\Omega_{\rm m}$);
(f) Einstein de Sitter. The blue bands indicate the $1\sigma$ (dark shade) and 
$2\sigma$ (light shade) confidence regions from figure~2.}
\label{figure5}
\end{figure}

\begin{figure}
\begin{center}
\includegraphics[width=4.3in]{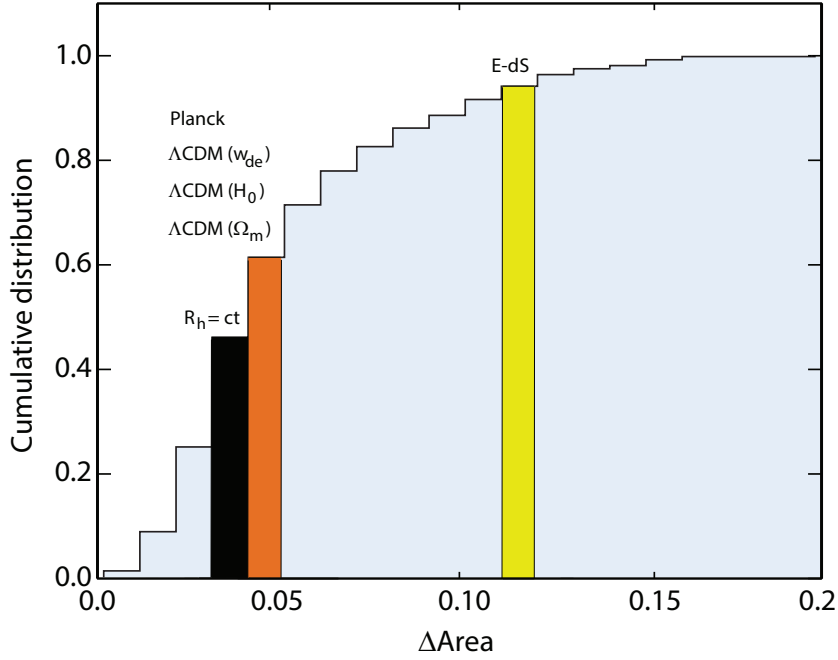}
\end{center}
\caption{Cumulative distribution (normalized to $1$) of the differential area
$\Delta A$ calculated for $H(z)$ according to Equation~(3.5) for mock samples
constructed via Gaussian randomization of the (modified) $H(z_i)$ values
shown in figure~2. The various models are shown according to their probabilities
listed in Table~2.}
\label{figure6}
\end{figure}

\begin{table*}
  \centerline{{\bf Table 2.} Model comparison based on the reconstructed $H(z)$}
  \centerline{with errors artificially reduced by $25\%$}\vskip 0.1in
\footnotesize
  \centering
  \begin{tabular}{lccccc}
&& \\
    \hline
\hline
&& \\
Model & $H_0$ & $\Omega_{\rm m}$ & $\Omega_{\rm de}$&$w_{\rm de}$&Prob. $(\%)$ \\
&(km s$^{-1}$ Mpc$^{-1}$)&&&& (Fig.~6) \\
&& \\
\hline
&& \\
$R_{\rm h}=ct$ & $62.7$& --& --& --& $54.7$ \\
$\Lambda$CDM ($w_{\rm de}$) & $67.4$& $0.314$ {\rm (fixed)}& $1-\Omega_{\rm m}$& $-0.94$& $39.6$ \\
$\Lambda$CDM ($H_0$)& $68.0$& $0.314$ {\rm (fixed)}& $1-\Omega_{\rm m}$& $-1\quad\;$& $39.6$ \\
$\Lambda$CDM ($\Omega_{\rm m}$)& $67.4$&\qquad $0.319$ {\rm (optimized)}& $1-\Omega_{\rm m}$& $-1\quad\;$& $39.6$ \\
Planck &$67.4$& $0.314$ {\rm (fixed)}& $1-\Omega_{\rm m}$& $-1\quad\;$& $15.2$ \\
E-dS &$42.0$& $1.0\quad$ {\rm (fixed)}& --& --& $7.3$ \\
&& \\
\hline\hline
  \end{tabular}
\end{table*}

\section{Conclusions}
We have used Gaussian Processes to reconstruct the Hubble constant $H(z)$ and its
associated $1\sigma$ and $2\sigma$ confidence regions, based exclusively on the
observation of cosmic chronometers. This is an important constraint, given that other
methods of measuring the Hubble constant invariably rely on the pre-assumption
of a particular cosmological model, which biases any subsequent analysis of the
data for the purpose of model selection. For example, using BAO to determine
the Hubble constant generally requires the assumption of a fiducial model in
order to separate the cosmological redshift of the BAO peak from effects
associated with internal redshift space distortions. This possible
contamination arises when the positions of galaxies are plotted in redshift-space 
rather than in terms of their angular-diameter distance, and is due to the
peculiar motion of galaxies subject to their mutual gravitational attraction, 
contributing a Doppler shift in addition to the cosmological redshift. In order to 
`measure' and subtract this Doppler effect, one must adopt a specific expansion scenario. 
Thus, the inferred BAO scale and $H(z)$ are relevant to that model, but not necessarily 
to the alternatives. And local measurements of $H(z)$, e.g., using Cepheid variables 
or other types of distance calibration, are often based on the 
parametrization in $\Lambda$CDM though, in principle, a re-calibration
could be made for each individual model. The reconstructed $H(z)$ used in 
this paper is completely model free, and is therefore suitable 
for probing the expansion history without any prejudice.

The reconstructed $H(z)$ confirms the results of earlier work \cite{7,8},
which showed that the Hubble constant measured with cosmic chronometers
does not support models with a variable expansion rate, preferring instead
the constant rate of expansion predicted by the $R_{\rm h}=ct$ universe. The
earlier results were based on optimizing model fits to the $H(z)$ data. The
$H(z)$ function reconstructed with the use of Gaussian Processes is,
in principle, arguably better because it is obtained without reference to
any model at all.  Even a simple comparison by eye of the reconstructed
$H(z)$ in figure~1 with the corresponding best fit curves in figure~1
of ref.~\cite{12} suggests that the cosmic chronometer data favour a
constant expansion rate over a variable one.

The key results of our analysis are summarized in Tables 1 and 2. Of
the six models we compared here, the one favoured by the reconstructed
$H(z)$---based on both the published measurements (figure~1) and (though
to a lesser extent) the data with artificially reduced dispersions
(figure~2)---is $R_{\rm h}=ct$. $\Lambda$CDM and its variations also
provide reasonably good fits---at least for $z\ll 2$ (see figs.~3 and
5)---but with smaller $p$-values. One of the least favoured models
is actually the version of $\Lambda$CDM with the {\it Planck}
optimization of parameters. It is also important to note that
the other versions of $\Lambda$CDM explored here were fitted 
using {\it Planck} priors on their free parameters, except for one
of them that is different in each case, while the $R_{\rm h}=ct$
model has only one free parameter ($H_0$), and no prior was applied
to it. Information criteria provide a greater penalty to less 
parsimonious models (i.e., those using a larger number
of free parameters), so a model comparison based, e.g., on the 
Bayes Information Criterion \cite{28}, would have produced an even 
greater disparity in probabilities than those listed in the tables.

The analysis reported in this paper continues to build the case for
the $R_{\rm h}=ct$ cosmology. The standard model accounts 
for the data at both high and low redshifts, but tension between its 
predictions and the observations grows as the error in the measurements 
drops to levels of a few percent. This is seen in the angular correlation
function of the CMB, in the BAO scale measured with the Ly-$\alpha$
cross-correlation function, and in the growth rate at $z<1$.
In purely technical terms, the chief difference between these two
models is that the former has a strictly constant equation of state,
corresponding to the zero active mass condition \cite{13,17},
while the latter has an equation of state that varies as the relative
abundances of the constituents in the cosmic fluid change with
redshift.

One would think that such a difference might produce only subtle changes
in the expansion rate and other observable signatures. That may be
true in some respects, but definitely not in others. For example, this
difference in the equation of state completely eliminates the horizon
problem, thereby obviating the need for an early phase of inflated
expansion \cite{29}. Considering how difficult it has been to
find a completely satisfactory model of inflation, the $R_{\rm h}=ct$
universe provides an acceptable alternative to the current 
inflationary $\Lambda$CDM paradigm. Such an outcome would 
bring to an end the 30-year endeavor to fix the horizon problem,
which emerged from the initially misunderstood expansion history 
of the Universe. The observations today suggest a cosmic age 
consistent with a constant expansion rate since the very beginning. 
The non-empty $R_{\rm h}=ct$ universe (as opposed to the empty, 
curvature-driven Milne universe) formally describes this
constantly-expanding cosmos, and very interestingly permits 
all areas of the Universe observable today to have reached 
thermal equilibrium well before the cosmic microwave background 
was produced. If correct, the $R_{\rm h}=ct$ describes a much
simpler, more elegant cosmos without the pathologies (e.g., the
horizon problem) requiring layers of `fixes' to address apparent
inconsistencies between the measurements at high and low redshifts. 

\acknowledgments
We are grateful to the anonymous referee for very helpful, detailed
comments that have led to a significant improvement in the presentation 
of our results. Some of this work was carried at the Instituto de Astrof\'isica 
de Canarias in Tenerife, and  was partially supported by grant 2012T1J0011 from 
The Chinese Academy of Sciences Visiting Professorships for Senior International 
Scientists.

\newpage
\begin{figure}
\begin{center}
\includegraphics[width=5.3in]{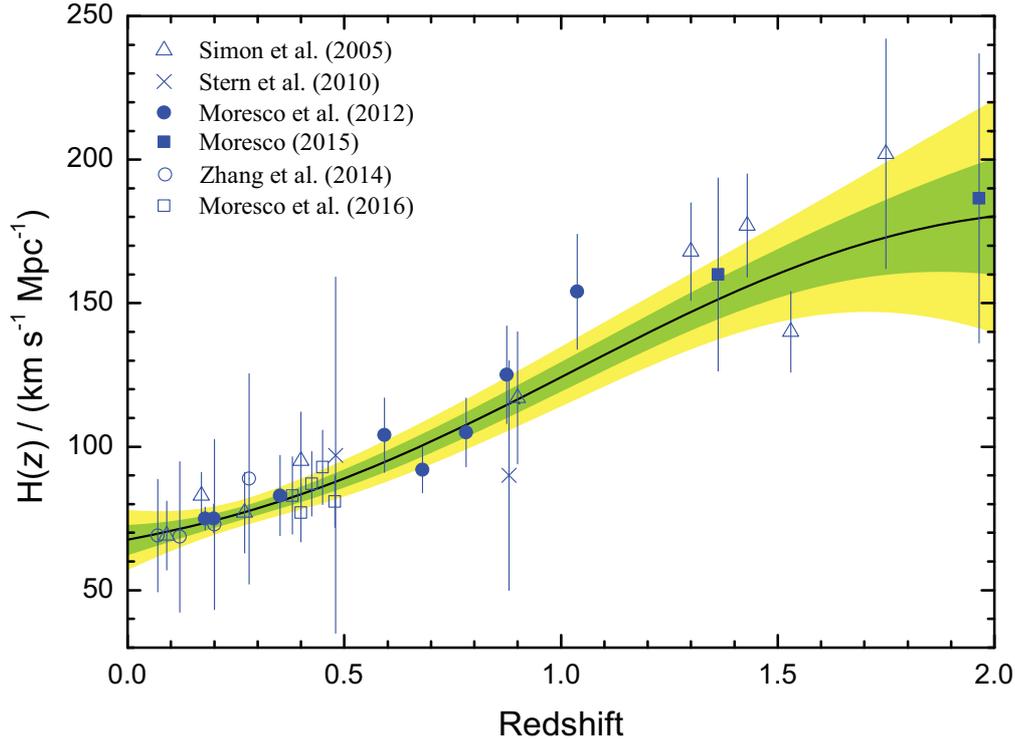}
\end{center}
\caption{The function $H(z)$ reconstructed with Gaussian Processes (solid black curve),
using the Matern92 covariance function (Seikel et al. 2012). The green and 
yellow shaded regions represent the $1\sigma$ and $2\sigma$ confidence regions
of the reconstruction, respectively.}
\label{figure7}
\end{figure}

\appendix
\section{Gaussian Processes with an Alternative Covariance Function}
The function $H(z)$ representing the 30 cosmic chronometer measurements 
shown in figure~1 is reconstructed here using a different covariance
function $k(x_1,x_2)$ than the `more conventional' one shown in 
Equation~(2.3). Specifically, we repeat the steps followed in producing
figs.~1, 3 and 4, but now with the Mat\'ern covariance function in
Gaussian processes, whose explicit form is \cite{5}
\begin{eqnarray}
k(x_1,x_2) &=& \sigma_f^2\exp\left(-{3|x_1-x_2|\over l}\right)\left(
1+{3|x_1-x_2|\over l}+{27|x_1-x_2|^2\over 7l^2}+\right. \nonumber \\
&\null&\qquad\qquad\qquad \left. {18|x_1-x_2|^3\over 7l^3}+{27|x_1-x_2|^4\over 35l^4}\right)\;.
\end{eqnarray}

The reconstructed function is shown in figure~7, along with the 30 
independent measurements of $H(z)$. This choice of covariance function
produces somewhat less smoothing than that using Equation~(2.3), with
the principal effect of creating greater tension between the predictions
of $\Lambda$CDM and the data at high redshift, where the standard model
requires a more rapid increase in $H(z)$ with $z$ than the reconstruction 
produces. The model comparisons are made in figure~8, and the corresponding 
cumulative probabilities are shown in figure~9 and tabulated in Table~3.

\begin{figure}[H]
\begin{center}
\begin{tabular}{cc}
\hskip-0.4in\includegraphics[width=0.62\linewidth]{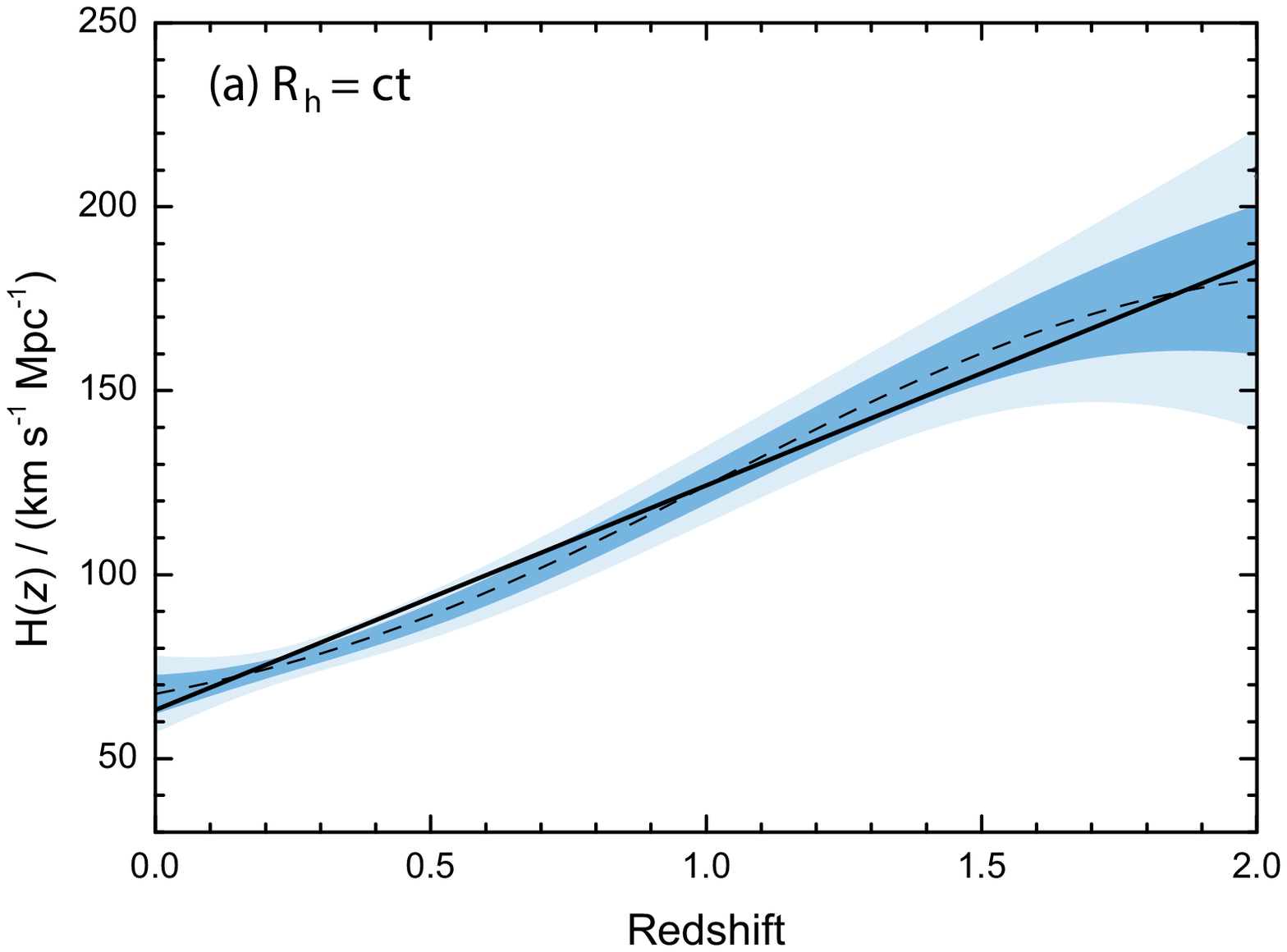} 
\hskip-0.6in\includegraphics[width=0.62\linewidth]{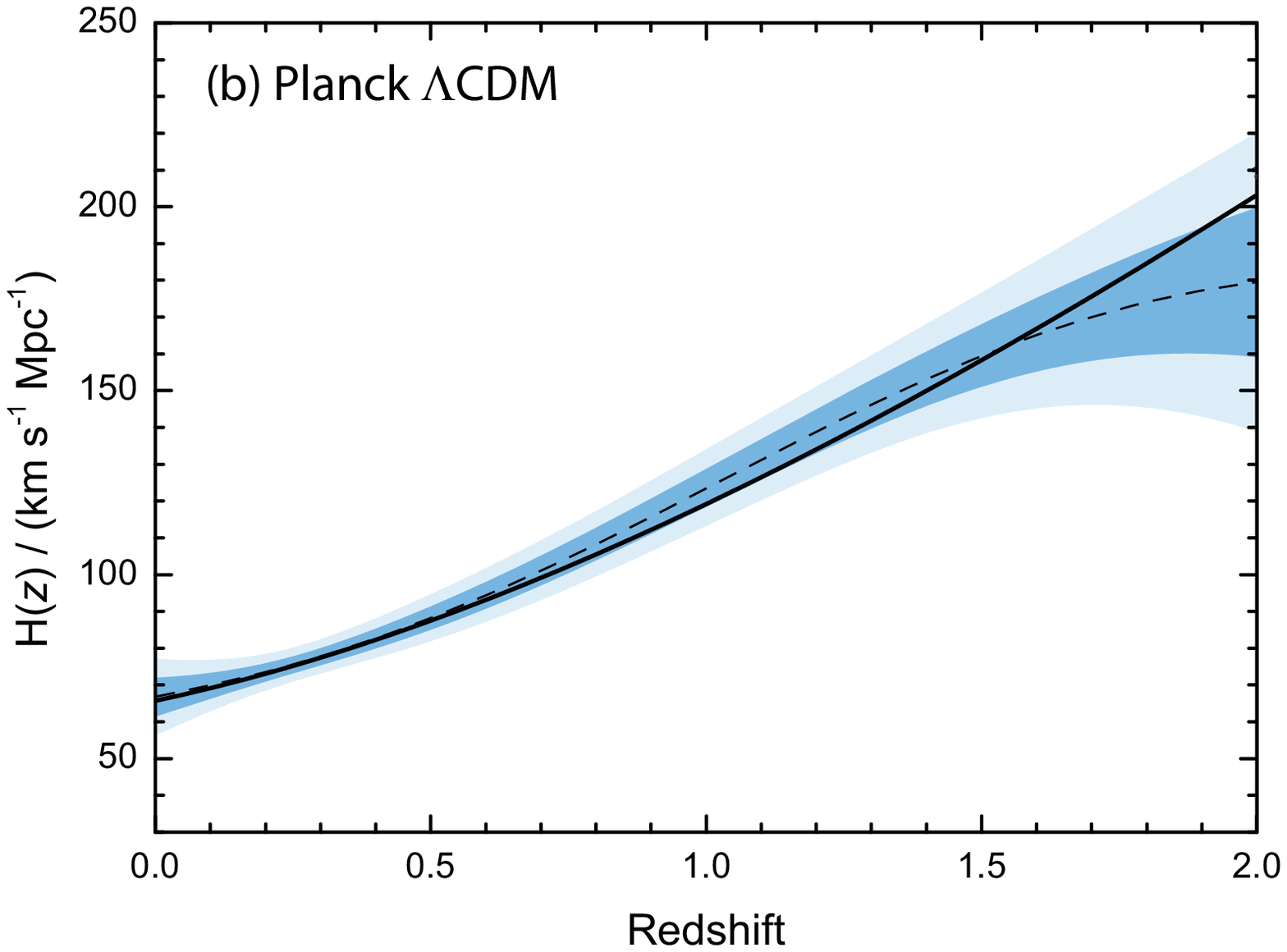} \\
\hskip-0.4in\includegraphics[width=0.62\linewidth]{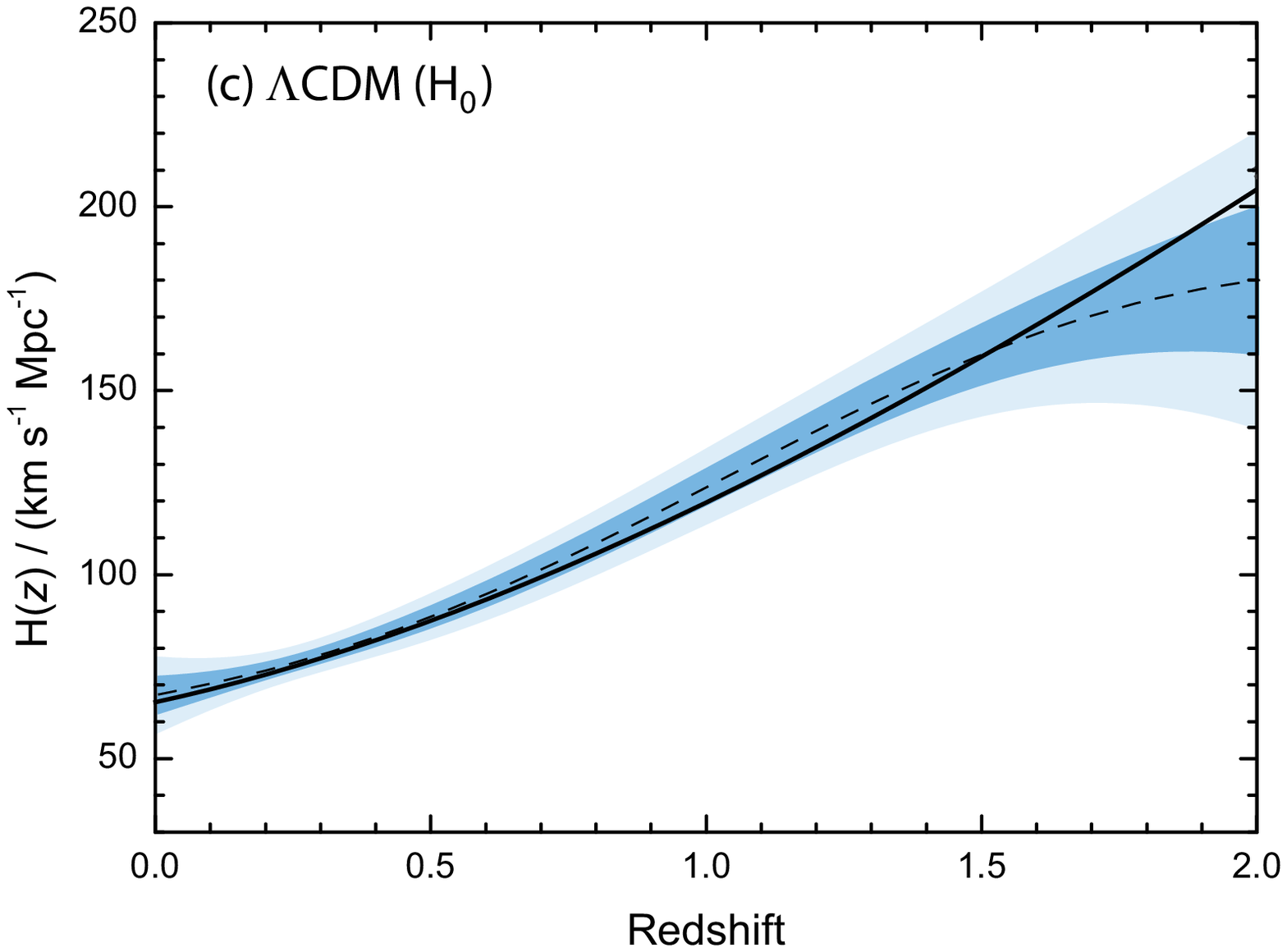}
\hskip-0.6in\includegraphics[width=0.62\linewidth]{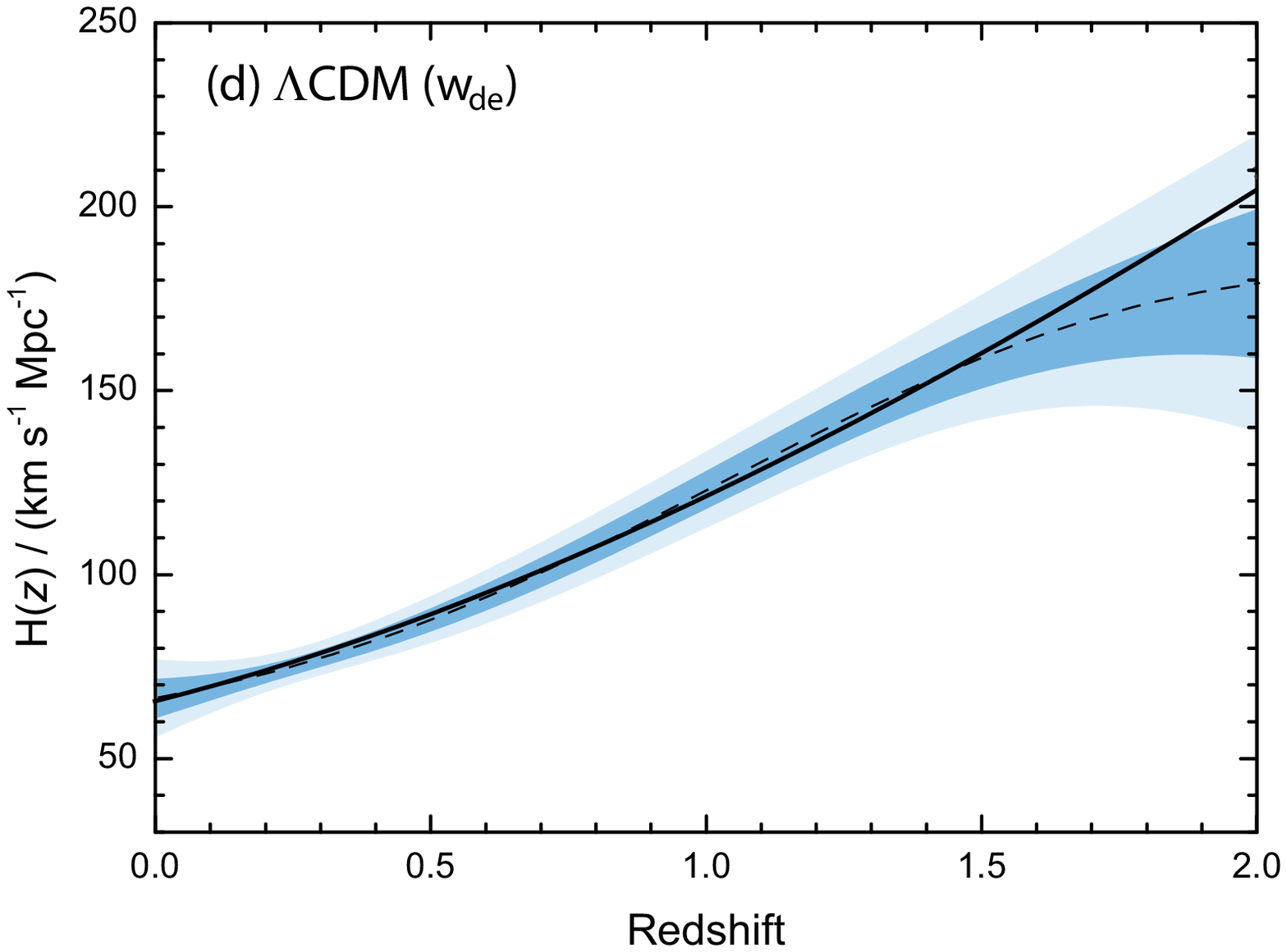} \\
\hskip-0.4in\includegraphics[width=0.62\linewidth]{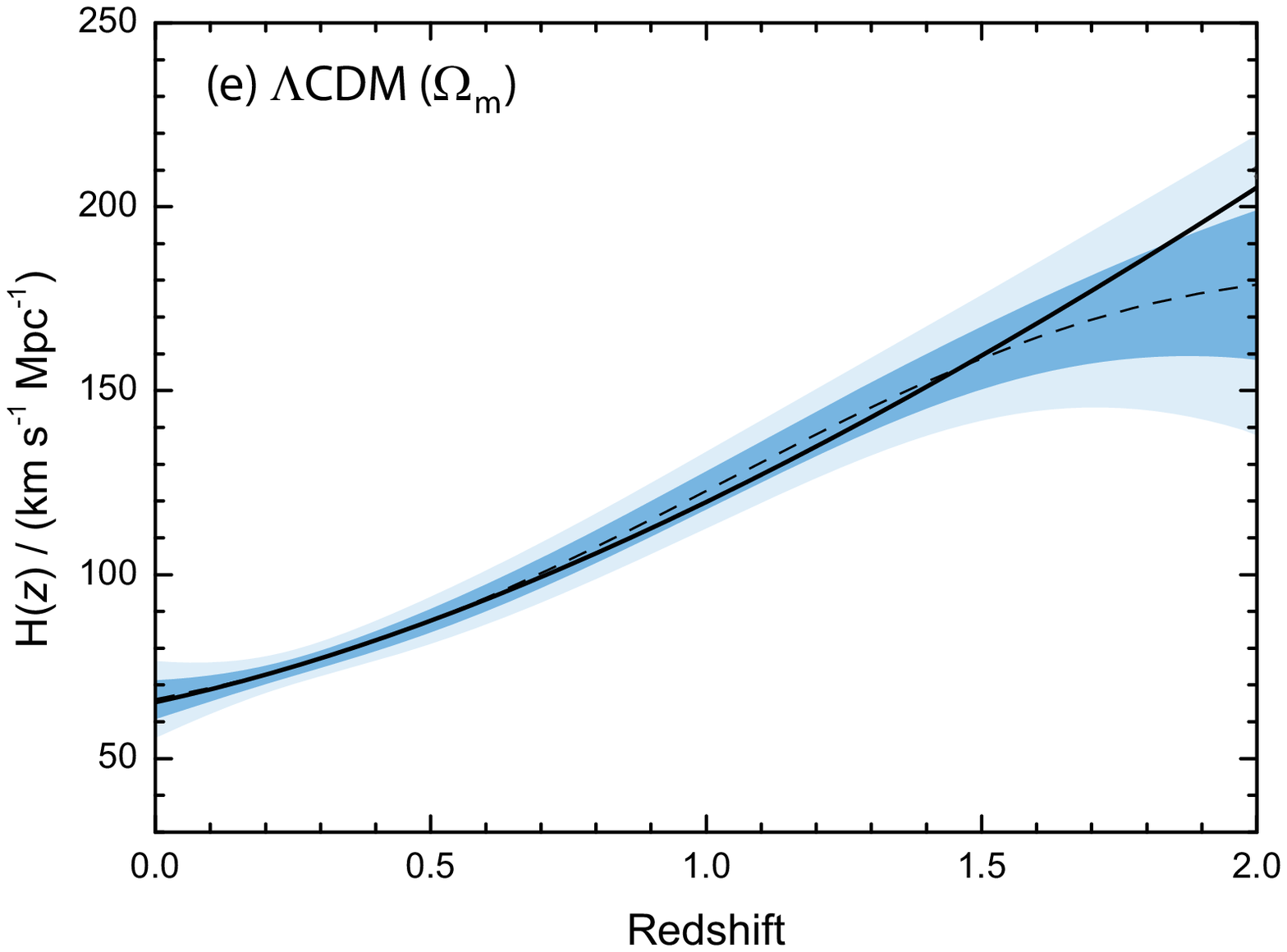}
\hskip-0.6in\includegraphics[width=0.62\linewidth]{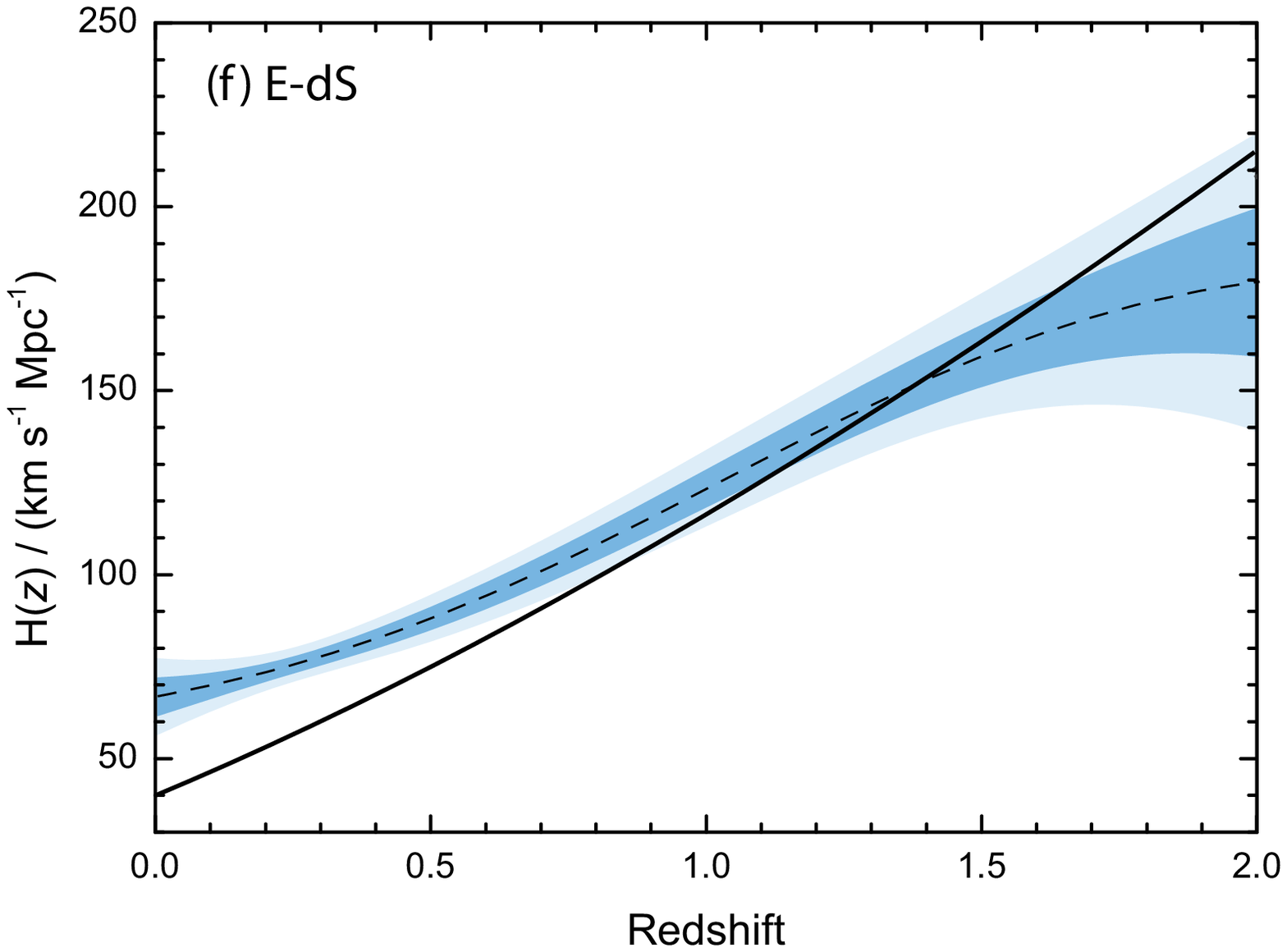}
\end{tabular}
\end{center}
\caption{The Hubble constant $H(z)$ (solid curve) in various cosmologies
optimized to fit the reconstructed function in figure~7 (dashed shade): (a)
The $R_{\rm h}=ct$ universe; (b) Planck $\Lambda$CDM; (c) $\Lambda$CDM with a prior, 
$\Omega_{\rm m}=0.314$; (d) flat $\Lambda$CDM, with prior density $\Omega_{\rm m}=0.314$ 
and Hubble constant $H_0=67.4$ km s$^{-1}$ Mpc$^{-1}$; and (e) $\Lambda$CDM ($\Omega_{\rm m}$);
(f) Einstein de Sitter. The blue bands indicate the $1\sigma$ (dark shade) and $2\sigma$ 
(light shade) confidence regions from figure~7. See Table~3 for re-optimized parameter values.}
\label{figure8}
\end{figure}

\begin{figure}
\begin{center}
\includegraphics[width=4.3in]{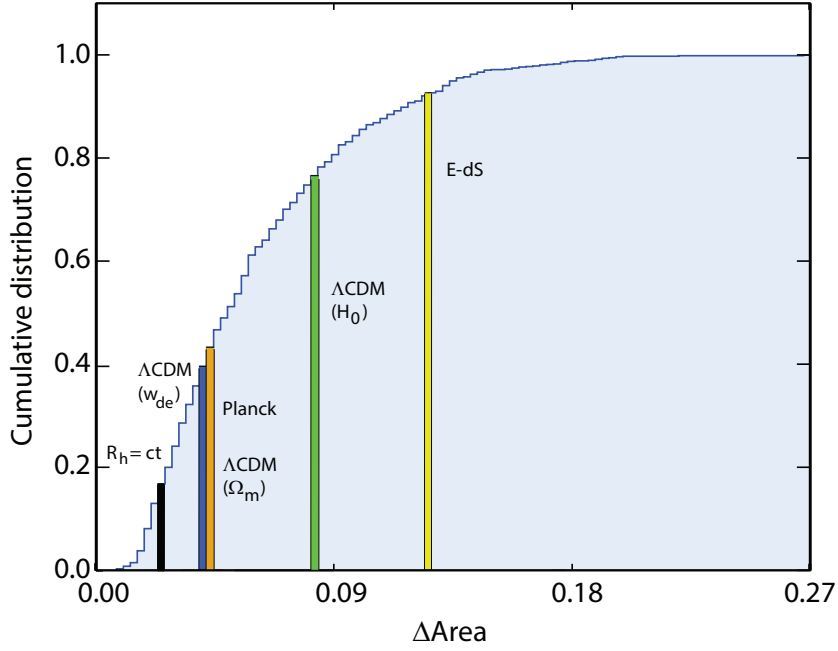}
\end{center}
\caption{Cumulative probability distribution (normalized to $1$) of the differential
area calculated for $H(z)$ according to Equation~(3.5) for mock samples
constructed via Gaussian randomization of the measured $H(z_i)$ values. This
calculation is based on the reconstruction shown in figs.~7 and 8, using the
Matern92 covariance function (Seikel et al. 2012). The various models 
are shown according to their probabilities listed in Table~3.}
\label{figure9}
\end{figure}

A comparison of Tables~1 and 3 shows that the choice of $k(x_1,x_2)$ in
Equation~(A.1) rather than Equation~(2.3) has resulted in slightly different
$p$-values for the various models, but their rank ordering has remained
the same. At least for cosmic chronometers, this choice of covariance
function in Gaussian Processes does not influence model selection in any
meaningful way.

\begin{table*}
\center
  \centerline{{\bf Table 3.} Model comparison using a reconstruction with the}
  \centerline{Matern92 covariance function (Seikel et al. 2012)}\vskip 0.1in
\footnotesize
  \centering
  \begin{tabular}{lccccc}
&& \\
    \hline
\hline
&& \\
Model & $H_0$ & $\Omega_{\rm m}$ & $\Omega_{\rm de}$&$w_{\rm de}$&Prob. $(\%)$ \\
&(km s$^{-1}$ Mpc$^{-1}$)&&&& (Fig.~9) \\
&& \\
\hline
&& \\
$R_{\rm h}=ct$ & $63.0$& --& --& --& $83.2$ \\
$\Lambda$CDM ($w_{\rm de}$) & $67.4$& $0.314$ {\rm (fixed)}& $1-\Omega_{\rm m}$& $-0.913$& $64.2$ \\
Planck $\Lambda$CDM&$67.4$ &$0.314$ {\rm (fixed)}&$1-\Omega_{\rm m}$&$-1$ &$60.4$ \\
$\Lambda$CDM ($\Omega_{\rm m}$)& $67.4$&\qquad $0.314$ {\rm (optimized)}& $1-\Omega_{\rm m}$& $-1$& $60.4$ \\
$\Lambda$CDM ($H_0$) &$67.8$& $0.314$ {\rm (fixed)}& $1-\Omega_{\rm m}$& $-1$& $23.4$ \\
E-dS &$41.3$& $1.0\quad$\;\;{\rm (fixed)}& --& --& $12.3$ \\
&& \\
\hline\hline
  \end{tabular}
\end{table*}


\begin{thebibliography}{99}
\bibitem[1]{1}Jimenez, R. \& Loeb, A., ``Constraining Cosmological Parameters Based on 
Relative Galaxy Ages," ApJ, \textbf{573}, 37 (2002)
\bibitem[2]{2}Moresco, M., Pozzetti, L., Cimatti, A., Jimenez, R., Maraston, C., 
Verde, L., Thomas, D., Citro, A., Tojeiro, R. \& Wilkinson, D., ``A $6\%$ measurement of 
the Hubble parameter at $z\sim 0.45$: direct evidence of the epoch of cosmic re-acceleration," 
JCAP, \textbf{05}, id 014 (2016)
\bibitem[3]{3}Rasmussen, C. \& Williams, C., ``Gaussian Processes for Machine Learning,"
Cambridge: The MIT Press (2006)
\bibitem[4]{4}Holsclaw, T., Alam, U., Sanso, B., Lee, H., Heitmann, K., 
Habib, S. \& Higdon, D., ``Nonparametric Dark Energy Reconstruction from Supernova Data," 
Phys. Rev. Lett., \textbf{105}, id 241302 (2010)
\bibitem[5]{5}Seikel, M., Clarkson, C. \& Smith, M., ``Reconstruction of dark energy and 
expansion dynamics using Gaussian processes," JCAP, {\bf 06}, 036S (2012)
\bibitem[6]{6}Moresco, M., Cimatti, A., Jimenez, R. et al., ``Improved constraints on the
expansion rate of the Universe up to $z \sim 1.1$ from the spectroscopic evolution
of cosmic chronometers," JCAP, \textbf{8}, id 006 (2012)
\bibitem[7]{7}Melia F. \& Maier, R., ``Cosmic Chronometers in the $R_{\rm h}=ct$ Universe," 
MNRAS, {\bf 432}, 2669 (2013)
\bibitem[8]{8}Melia, F. \& McClintock, T. M., ``A Test of Cosmological Models Using
High-$z$ Measurements of $H(z)$," AJ, \textbf{150}, id 119 (2015)
\bibitem[9]{9}Melia, F., ``The Cosmic Horizon," MNRAS, {\bf 382}, 1917 (2007)
\bibitem[10]{10}Melia, F. \& Abdelqader, M., ``The Cosmological Spacetime,"  {\bf IJMP-D}, 18, 1889 (2009)
\bibitem[11]{11}Melia, F. \& Shevchuk, A., ``The $R_{\rm h}=ct$ Universe," MNRAS, {\bf 419}, 2579 (2012)
\bibitem[12]{12}Wei, J.-J., Melia, F. \& Wu, X.-F., ``Impact of a Locally Measured $H_0$ on
the Interpretation of Cosmic Chronometer Data," ApJ, \textbf{835}, 270 (2017)
\bibitem[13]{13}Shi X., ``Peculiar Hubble Flows in Our Local Universe," ApJ, \textbf{486}, 32 (1997)
\bibitem[14]{14}Keenan, R. C., Barger, A. J. \& Cowie, L. L., ``Evidence for a $\sim 300$ Megaparsec 
Scale Under-density in the Local Galaxy Distribution," ApJ, \textbf{775}, 62 (2013)
\bibitem[15]{15}Romano, A. E., ``Hubble trouble or Hubble bubble?" e-print (arXiv:1609.04081) (2017)
\bibitem[16]{16}Melia, F., ``Physical Basis for the Symmetries in the 
Friedmann-Robertson-Walker Metric," Frontiers of Physics, {\bf 11}, 119801 (2016)
\bibitem[17]{17}Melia, F.,``The Zero Active Mass Condition in Friedmann-Robertson-Walker Cosmologies," 
Front Phys, {\bf 12}, 129802 (2017)
\bibitem[18]{18}Blake, C. et~al., ``The WiggleZ Dark Energy Survey: joint measurements of
the expansion and growth history at $z < 1$," MNRAS, \textbf{425}, 405 (2012)
\bibitem[19]{19}Treu, T., Ellis, R. S., Liao, T. X., van Dokkum, P. G., Tozzi, P., 
Coil, A., Newman, J., Cooper, M. C. \& Davis, M., ``The Assembly History of Field Spheroidals: Evolution of
Mass-to-Light Ratios and Signatures of Recent Star Formation," ApJ, \textbf{633}, 174 (2005)
\bibitem[20]{20}Simon, J., Verde, L. \& Jimenez, R., ``Constraints on the redshift dependence
of the dark energy potential," PRD, \textbf{71}, id 123001 (2005)
\bibitem[21]{21}Stern, D., Jimenez, R., Verde, L., Stanford, S. A. \& Kamionkowski, M., ``Cosmic
Chronometers: Constraining the Equation of State of Dark Energy. II. A Spectroscopic
Catalog of Red Galaxies in Galaxy Clusters," ApJS, \textbf{188}, id 280 (2010)
\bibitem[22]{22}Zhang, C., Zhang, H., Yuan, S., Liu, S., Zhang, T.-J., Sun \& Y.-C., ``Four new observational
$H(z)$ data from luminous red galaxies in the Sloan Digital Sky Survey data release seven,"
Res. Astron. Astrophys., \textbf{14}, 1221 (2014)
\bibitem[23]{23}Moresco, M., ``Raising the bar: new constraints on the Hubble parameter
with cosmic chronometers at $z\sim 2$," MNRAS, \textbf{450}, L16 (2015)
\bibitem[24]{24}Leaf, K. \& Melia, F., ``Analyzing H(z) data using two-point diagnostics,"
MNRAS, \textbf{470}, 2320 (2017)
\bibitem[25]{25}L\'opez-Corredoira, M., Vazdekis, A., Guti\'errez, C. M. \& Castro-Rodr\'iguez, N.,
{A\&A}, in press (arXiv:1702.00380) (2017)
\bibitem[26]{26}Planck Collaboration, ``Planck 2013 results. XXIII. Isotropy and statistics of the CMB,"
A\&A, \textbf{571}, id A23 (2014).
\bibitem[27]{27}Yennapureddy, M. K. and Melia F., ``Reconstruction of the HII Galaxy Hubble Diagram 
using Gaussian Processes," JCAP, \textbf{ JCAP11(2017)029}, 029 (2017)
\bibitem[28]{28}Schwarz, G., ``Estimating the Dimension of a Model," Ann. Statist., \textbf{6}, 461 (1978)
\bibitem[29]{29}Melia, F., ``The $R_{\rm h}=ct$ universe without inflation," A\&A, \textbf{553}, id A76 (2013)

\end{thebibliography}
\end{document}